\documentclass[manuscript]{emulateapj}
\usepackage{amsmath}

\def\plotfiddle#1#2#3#4#5#6#7{\centering \leavevmode
    \vbox to#2{\rule{0pt}{#2}}
    \includegraphics{#1}}

\newcommand{\Mstar}{M_*}

\newcommand{\Msun}{M_{\odot}}

\newcommand{\Rstar}{R_*}
\newcommand{\Rsun}{R_\odot}
\newcommand{\Msunperyr}{M_{\odot}\,{\rm yr}^{-1}}

\newcommand{\Lx}{L_{\rm X}}
\newcommand{\Tx}{T_{\rm X}}
\newcommand{\nH}{n_{\rm H}}

\newcommand{\percc}{\rm \,cm^{-3}}
\newcommand{\psqcm}{{\rm cm}^{-2}}
\newcommand{\persqcm}{\rm \,cm^{-2}}

\newcommand{\pers}{{\rm s}^{-1}}
\newcommand{\pyr}{{\rm yr}^{-1}}
\newcommand{\ccps}{{\rm cm}^{3} {\rm s}^{-1}}

\newcommand{\ergpers}{{\rm erg\,s^{-1}}}

\newcommand{\NH}{N_{\rm H}}
\newcommand{\Htwo}{{\rm H_{2}}}
\newcommand{\Ntwo}{{\rm N_{2}}}
\newcommand{\Otwo}{{\rm O_{2}}}
\newcommand{\HtwoO}{{\rm H_2O}}
\newcommand{\CtwoHtwo}{{\rm C_2H_2}}

\newcommand{\COtwo}{{\rm CO_2}}

\newcommand{\CHfour}{{\rm CH_4}}
\newcommand{\NHthree}{{\rm NH_3}}
\def\micron{\hbox{$\,\mu$m}}

\newcommand{\Hp}{{\rm H}^+}
\newcommand{\Hep}{{\rm He}^+}
\newcommand{\SOtwo}{{\rm SO_{2}}}
\def\micron{\hbox{$\,\mu$m}}
\newcommand{\be}{\begin{equation}}
\newcommand{\ee}{\end{equation}}

\begin{document}


\title{Formation of Organic Molecules and Water in Warm Disk Atmospheres} 


\author{Joan R.\ Najita}
\affil{National Optical Astronomy Observatory, 950 N. Cherry Avenue, Tucson, AZ, 85719}

\author{M\'at\'e \'Ad\'amkovics and Alfred E.\ Glassgold}
\affil{Astronomy Department, University of California, Berkeley, CA 94720}



\begin{abstract}
Observations from {\it Spitzer} and ground-based infrared spectroscopy
reveal significant diversity in the molecular emission 
from the inner few AU of T Tauri disks.
We explore theoretically the possible origin of this diversity by
expanding on our earlier thermal-chemical model of disk atmospheres.
We consider how variations in grain settling, X-ray irradiation,
accretion-related mechanical heating, and the oxygen-to-carbon ratio
can affect the thermal and chemical properties of the atmosphere at 
0.25--40\,AU.
We find that these model parameters can account for
many properties of the detected molecular emission.  The
column density of the warm (200-2000K) molecular atmosphere is
sensitive to grain settling and the efficiency of accretion-related
heating, which may account, at least in part, for the large range
in molecular emission fluxes that have been observed.  The dependence
of the atmospheric properties on the model parameters may also help
to explain trends that have been reported in the
literature between molecular emission strength and mid-infrared
color, stellar accretion rate, and disk mass.
We discuss whether
some of the differences between our model results and the observations 
(e.g., for water) indicate a role for vertical transport and
freeze-out in the disk midplane.  We also discuss how planetesimal
formation in the outer disk (beyond the snowline) may imprint a
chemical signature on the inner few AU of the disk and speculate
on possible observational tracers of this process.  
\end{abstract}






\section{Introduction}

Recent work with the {\it Spitzer Space Telescope} has found that
emission from water and organic molecules is common in spectra of
T Tauri stars (Carr \& Najita 2011, 2008; Salyk et al.\ 2011, 2008; Pontoppidan
et al.\ 2010; Pascucci et al.\ 2009), with the emission likely arising
from the inner few AU region of the circumstellar disk (Carr \&
Najita 2008; Salyk et al.\ 2008).  These new molecular diagnostics
complement the more familiar near-infrared molecular diagnostics
(CO, water, OH) that probe disk radii $\lesssim 1$\,AU (see 
Najita et al.\ 2007; Carmona 2010 for recent reviews).
Since T Tauri disks are optically thick in the continuum, these
molecular emission features presumably arise from a temperature
inversion region in a warm disk atmosphere.  The possibility of
warm molecular emission from the atmospheres of inner disks is
supported by thermal-chemical models of disk atmospheres (Glassgold
\& Najita 2001; Glassgold et al.\ 2004; Kamp \& Dullemond 2004; 
Willacy \& Woods 2009; Nomura \& Millar 2005; Gorti \& Hollenbach 2008). 

One of the interesting properties of the mid-infrared molecular
emission observed with {\it Spitzer} is that the fluxes and relative 
strengths of molecular features vary among the T Tauri stars 
(Carr \& Najita 2011; 
Salyk et al.\ 20011; 
Pontoppidan et al.\ 2010; 
Pascucci et al.\ 2009). 
The strength of the water emission varies by a factor of $> 30$ 
(e.g., Pontoppidan et al.\ 2010), and  
an equally large range is found for the strength of the HCN emission 
(e.g., Salyk et al.\ 2011). 
The water emission can be brighter or fainter than the HCN emission,  
and HCN is usually stronger than $\CtwoHtwo$ among T Tauri stars but is 
sometimes weaker (Carr \& Najita 2011; Salyk et al.\ 2011; 
Pontoppidan et al.\ 2010; Pascucci et al.\ 2009).  
What is the origin of these variations in flux and relative strength?

Several interesting possibilities are suggested by trends in the
demographics of molecular emission sources.  While HCN is typically
stronger than $\CtwoHtwo$ in T Tauri stars, Pascucci et al.\ (2009) found
$\CtwoHtwo$ to be stronger than HCN in median spectra of brown dwarfs.  Pascucci
et al.\ suggested that this dichotomy arises because of the stronger
UV field of T Tauri stars, which enhances the dissociation of $\Ntwo$ and the
formation of HCN in those systems.

Further possible trends in HCN emission strength were noted by 
Teske et al.\ (2010).  They found that the HCN emission strength 
of T Tauri stars 
appears to increase with stellar accretion rate and
with X-ray luminosity, suggesting that these parameters may play a role in the
molecular emission fluxes.  They suggested that the trend of HCN flux with
stellar accretion rate might result from enhanced accretion-related
mechanical heating in the disk atmosphere, along the lines explored
theoretically by Glassgold et al.\ (2009).  
These authors had
previously suggested that increased rates of accretion-related
mechanical heating can increase the column density of warm water
in the T Tauri disk atmosphere, producing stronger water emission.  
A similar effect might also be expected for other molecular species. 
Consistent with this perspective, 
Salyk et al.\ (2011) have reported a higher detection rate of molecular 
emission in systems with larger H$\alpha$ equivalent widths 
(and tentatively also with stellar accretion rate), which 
they interpret as suggesting that accretional heating 
plays a role in enhancing the line emission. 
Salyk et al.\ also find that molecular emission is preferentially 
detected from disks that have experienced more grain 
settling, as indicated by their mid-infrared colors. 

As an alternate explanation for their results, 
Teske et al.\ also discussed the possibility 
that increased UV irradiation produced at higher stellar accretion 
rates might enhance the HCN abundance, as discussed by Pascucci 
et al.\ (2009).  Little interpretation was offered for the 
possible trend they found with $\Lx$, 
other than that X-ray irradiation may help  
to synthesize HCN by raising the abundance of molecular ions and 
radicals.

Carr \& Najita (2011) also reported a possible relation 
between the HCN/$\HtwoO$ emission flux ratio 
and disk mass in their small sample of T Tauri star spectra.  
The sources with the largest HCN/$\HtwoO$ flux ratios are those 
with the largest disk masses. 
They note that such a trend might be expected as a consequence of 
planetesimal and protoplanet formation in the outer disk.  
As described by Ciesla \& Cuzzi (2006 and references therein),
because icy bodies ranging from centimeter to kilometer and larger 
size migrate at different rates, their migration can lead
to enhancements or depletions of water vapor in the disk that may
vary as a function of both disk radius and time.  
Since water is a significant reservoir of oxygen, 
these effects can lower or raise the gas-phase oxygen-to-carbon 
(O/C) ratio in the 
inner disk and possibly affect the HCN and $\HtwoO$ abundances in the 
manner observed. 
More generally, because it is difficult to diagnose the formation 
of kilometer-sized planetesimals, the building blocks of planets, 
by other means, it is interesting to explore whether their formation 
may induce this (or any other) chemical signature in the inner disk 
that may signal their formation.

Here we explore the above ideas by
expanding on our earlier thermal-chemical model of disk atmospheres.
Models of disk atmospheres are in an early stage of development.  In
part, this is because disks are complex systems, and many processes
potentially play a role in determining the thermal-chemical properties
of their atmospheres.  The relevant processes include those familiar
from the study of molecular clouds and photodissociation regions
(e.g., UV and X-ray irradiation and their roles in heating,
photochemistry, and ionization; neutral and ion-molecule reactions;
and the freeze out and desorption of molecules on grains).  Other
potentially important processes are more unique to circumstellar
disks and are generally less well understood, both in terms of how
they operate and their impact on the thermal-chemical properties
of disks. These include grain growth and settling, mechanical heating
related to the magnetorotational instability, and radial and/or
vertical transport of gas and solids.  Models are developed
to gain insight into how these processes affect disk atmospheres,
with earlier studies having concentrated on different sets of
processes and the interplay between them.

In our previous work, we constructed a model of inner disk atmospheres 
that includes the thermal decoupling of the gas and dust 
(e.g., Glassgold et al.\ 2001, 2004; see also Kamp \& Dullemond 2004;
Nomura \& Millar 2005), was based on X-ray heating and ionization, and 
included effects such as  
grain growth and accretion-related mechanical heating (Glassgold et al.\ 2004). 
We used this model to study the gas-phase chemistry of water, 
including the effect of $\Htwo$ formation on warm grain surfaces 
(Glassgold et al.\ 2009). 
Other studies have emphasized the effect of UV irradiation on the 
disk (e.g., Nomura \& Millar 2005; Woitke et al.\ 2009; Woods \& Willacy 2009; 
Bethell \& Bergin 2009),
included freeze-out and desorption in the disk midplane 
or outer disk (e.g., Markwick et al.\ 2002;
Woods \& Willacy 2009; Walsh et al.\ 2010), 
and explored the effect of transport processes 
on disk chemistry (e.g., Ilgner et al.\ 2004, Semenov et al.\ 2006; 
Willacy et al.\ 2006; Heinzeller et al. 2011), processes that we have yet to include.   
In this paper, we build on our earlier work, implementing an improved 
X-ray ionization theory and an expanded chemical network and explore how 
variations in X-ray irradiation, accretion-related mechanical 
heating, and the O/C ratio affect the thermal and chemical 
properties of the atmosphere.

Our study complements other recent work on the chemistry of inner disks
(e.g., 
Ag\'undez et al.\ 2008; 
Woitke et al.\ 2009; 
Willacy \& Woods 2009; 
Nomura et al.\ 2009; 
Kress et al.\ 2010;
Markwick et al.\ 2002). 
The work of Ag\'undez et al.\ (2008) is perhaps the closest in spirit 
to our own study.  In their study, they addressed the steady-state 
properties of the 
disk atmosphere, assuming that the disk is strongly irradiated by FUV photons
and adopting an extended neutral chemistry network that includes carbon species.  
Our study complements the Ag\'undez et al.\ study in that we include 
X-ray (but not UV) irradiation of the disk.  
While a more complete model would include UV irradiation, 
here we explore how well a model that excludes UV can do in accounting 
for the existing observations.

\section{Description of the Model}
We study the hydrocarbon chemistry of protoplanetary disks with the
steady-state thermal-chemical model developed by Glassgold, Najita
\& Igea (2004; henceforth GNI04). This model is based on the D'Alessio
et al.~(1999) disk atmosphere for a generic T Tauri star of mass
$0.5\Mstar$,  blackbody temperature $T_* = 4000$\,K, and accretion
rate $10^{-8} \Msun \pyr$.  The disk is irradiated by stellar X-rays 
(which affect the temperature, ionization, and chemical  structure
of the gas), but not by far ultraviolet (FUV) radiation. GNI04 used
the dust temperature obtained by D'Alessio et al.\ and calculated
the gas temperature on the basis of heating by collisions with dust
grains, X-rays, and accretion-related mechanical heating, balanced
by a variety of line cooling processes and gas-dust collisions. 
As a result, the gas temperature is elevated over the dust temperature 
over much of the atmosphere. 

We adopt a steady state treatment of the chemical kinetics rather 
than a time-dependent one. 
This approach is supported in part by the warm gas temperatures 
inferred for the {\it Spitzer} molecular emission (300--1000\,K), 
which generally imply short chemical timescales 
compared to transport timescales.
The chemistry in GNI04 was limited to 25 species for
the elements H, He, C and O, and it contained 115 reactions.  In a
subsequent study of the water chemistry of protoplanetary disks,
Glassgold, Meijerink \& Najita (2009; henceforth GMN09) corrected
and expanded the chemistry to $\sim 100$ species and $\sim 400$
reactions. 

GMN09 emphasized the role of $\Htwo$ formation on warm dust
grains, following the theory of Cazaux \& Tielens  (2002, 2004;
Cazaux et al.~2005) who showed that H$_2$ can form with moderate
efficiency ($ \lesssim 0.2$) on warm dust with temperatures up to
$\sim 900$\,K.
Other recent studies of disk atmospheres have also employed this 
process in their models (e.g., Heinzeller et al.\ 2011; 
Woitke et al.\ 2009; Willacy \& Woods 2009; and Woods \& Willacy 2009).
Enhanced $\Htwo$ formation on grains appears to play an important role in producing 
significant warm molecular columns in models such as ours that include 
grain growth (Woods \& Willacy 2009). 
If grain growth is not included, enhanced $\Htwo$ formation on grains can lead 
to the overproduction of warm molecular emission (e.g., Heinzeller et al.\ 2011).

In this paper we build on our previous models 
by employing an expanded chemical network that includes hydrocarbon,
nitrogen and sulfur chemistry. The current model includes an enhanced theory
of X-ray ionization that treats individually the nine most abundant
heavy atoms (Adamkovics, Glassgold, \& Meijerink 2011; henceforth 
AGM11) in addition
to C, N and O. The chemical network now contains $\sim 115$ species
and $\sim 1200$ reactions. Even so, the molecular complexity is
limited to molecules (and ions) of 6 atoms or less, with hydrocarbons
containing no more than two carbon atoms, as in C$_2$H$_4$. 

The reaction list contains neutral and ion-molecule reactions, X-ray
interactions, dissociative and radiative recombination, and radiative
association. We evaluated the reactions in our model using the
primary literature and the critical evaluations of Baulch et al.\
(2005) for neutral reactions and Anicich (1993) for ion-molecule
reactions. We have also consulted the NIST Chemical Kinetics Database
(http://kinetics.nist.gov) as well as the UMIST (Woodall et al.
2007) and KIDA (Wakelam, 2009) collections.

Irradiation by stellar and/or interstellar FUV is not included in
the present model, an assumption that would apply when the disk is well
shielded from UV (e.g., by an intervening funnel flow).  This
assumption may also be more generally valid in the context of
heating.  Earlier work (e.g., Nomura et al.\ 2007) has shown that 
UV heating due to the photoelectric effect is significantly reduced 
relative to other heating processes when grains have grown and
settled to the extent that has been deduced for T Tauri disks
(Furlan et al.\ 2006). This, and the
uncertain PAH abundance of T Tauri disks, mitigates our neglect of UV
heating, a topic that we return to in section 4. More seriously,
UV photodissociation likely affects the molecular abundances as we
also discuss in section 4. 

One obvious difference between our X-ray irradiated model and others that 
focus on UV irradiation is that in our model molecules are destroyed by H$^+$ 
and He$^+$. In the presence of UV irradiation, photodissociation would also 
play a significant role in destroying molecules.
In a similar vein, UV irradiation will also alter the ionization
and temperature structure from that considered here.  
Another difference between UV and X-ray irradiated models
is that
X-rays are also able to ionize species that are unavailable to UV photons. 
For example, although FUV photons are unable to directly ionize N and $\Ntwo$, 
ion-neutral reactions of $\Ntwo$ with H$_3^+$ and He$^+$ (which are generated by 
X-rays) can activate the nitrogen chemistry. 

We include the formation of $\Htwo$ on dust
grains in the same way as GMN09 (their eqs.\ 14 and 15), following the
theory of Cazaux \& Tielens. At the suggestion of Bai \& Goodman
(2009), Cazaux \& Tielens (2010) have since corrected their
calculations, but the changes are only of a quantitative nature.
We ignore the possibility of forming other molecules on grains.
GMN09 found that ignoring $\HtwoO$ formation on grains did not much
affect the distribution of gaseous $\HtwoO$. We also ignore cooling
by $\HtwoO$ rotational transitions as well as heating 
due to UV absorption by $\HtwoO$.  Water line emission has been measured to
be a significant coolant for disk atmospheres (Pontoppidan et al.\
2010).  These limitations will be remedied in future reports.

As discussed in more detail in AGM11, the abundances of the heavy
elements play an important role in the chemistry of the disk
atmosphere, because they are the dominant charge carriers below the
atomic to molecular transition. Measurements are not available for
the absolute gas-phase abundances of the elements in thick molecular
clouds, not to mention protoplanetary disks.  The only abundance determinations
come from UV absorption line spectroscopy of relatively diffuse
interstellar clouds ($A_V \lessapprox 2$\, mag) obtained from
observations made with satellite observatories.  The measured
abundances of the heavy elements vary from cloud to cloud, and 
we continue to use those abundances reported for 
the very well studied cloud towards $\zeta$ Oph
(Savage \& Sembach 1996); as shown in Table 1. The last column shows
the depletion factors, defined as the ratio of solar to interstellar
abundances, with the former coming from Table 1 of the review of
solar abundances by Asplund et al.\ (2009). As discussed in this
paper and in Jenkin's (2009)  review of interstellar abundances,
both sets of values have significant uncertainties, in some cases factors of
a few. Jenkins highlights the case of S, where the UV observations
suggest essentially no depletion in interstellar clouds, despite
the fact that this element is seen in meteorites.

Although we use the depleted interstellar abundances in Table 1 for
the preliminary modeling purposes of this paper, we recognize that
the abundances of the heavy elements in diffuse interstellar clouds
may well have been altered in dense clouds and by the formation and
evolution of protoplanetary disks. The determination of relative
abundances in disks remains a primary goal of observations of the
emission line fluxes of their atoms and molecules. As discussed
in Section 1, we will examine  one particular modification of the
elemental abundances, the variation of the oxygen-to-carbon
ratio (O/C) that may be induced by the adsorption of water 
on solid bodies and the transport
of those bodies between large and small radii.

The molecular abundances in the inner disk atmosphere are affected
by the level of X-ray irradiation, the strength of the accretion-related
mechanical heating, and the oxygen-to-carbon ratio (O/C). Of course they
also change with the radial distance from the star. We have
investigated these effects by carrying out variational calculations
about a ``reference case'' defined in Table 2. The X-ray variations
are parameterized by the X-ray luminosity $\Lx$ and the X-ray
temperature $T_{\rm X}$, and the mechanical heating by the parameter
$\alpha_h$ in Eq.~(12) in GNI04,
\begin{equation}
\label{accheat1}
\Gamma_{\rm acc} = \frac{9}{4} \alpha_h \rho c^2 \Omega,
\end{equation}	
where $\rho$ is the local mass density, $c$ is the isothermal sound
speed, $\Omega$ is the angular rotation speed and $\alpha_h$ is a
phenomenological parameter specifying the strength of the heating.
As discussed by GMN09 in connection with their Equation 1, the grain
surface area per unit volume is proportional to the dust-to-gas
ratio of the disk, $\rho_{\rm {dust}} / \rho$, and the inverse of the  
geometric mean ($a_g$) 
of the minimum and the maximum grain radii in an MRN dust size
distribution ($n(a)\propto a^{-3.5}$). 

Table 2 gives the range of parameter variations we considered. 
As in our previous papers, the dust-to-gas ratio was fixed at 0.01, 
and we considered values of $a_g$ that are 32 and 320
times larger than the usual interstellar value of $0.0224\,\micron$,
i.e., 0.707 and $7.07\,\micron$.  
This range is consistent with the dust depletion factors ($\sim$100--1000) 
that are found for the upper disk layers of T Tauri stars based on 
their mid-infrared colors (Furlan et al.\ 2006). 
We adopted $\Lx$ and $T_{\rm X}$ values that are typical of T Tauri 
stars ($L_X= 10^{29}-10^{31}\,\ergpers$; Feigelson et al.\ 2005) 
as well as lower values of $\Lx$ down to $10^{26}\,\ergpers$ 
in order to illustrate the role of the X-rays.  

The parameter $\alpha_h$ was varied so that accretion heating was either 
non-dominant ($\alpha_h=0.01$) or dominant ($\alpha_h=1$) as a heat source.  
The lower value is motivated by the viscosity parameter for 
T Tauri disks, which has been estimated as $\alpha \simeq 0.01$ 
as a vertical average in a steady state disk (e.g., Hartmann et al.\ 1998).  
The larger value is motivated by 3D MHD calculations that suggest that turbulent
accretion disks driven by the magnetorotational instability experience
greater energy dissipation at the disk surface than in a classical
``$\alpha$-disk''.  A recent calculation by Hirose \& Turner (2011)
suggests that the energy dissipation rate rises toward the disk
surface, reaching values that correspond to $\alpha_h\simeq 1$.

The range of radii we considered covers the inner disk region probed by {\it Spitzer} 
(the inner few AU) and the larger radii relevant to {\it Herschel} (out to 40 AU).
We considered a range in the O/C ratio in order 
to explore the dehydrated to super-hydrated regimes described in 
Ciesla \& Cuzzi (2006).  Those models indicate that the 
(vertically averaged) water vapor concentration can 
increase by up to an order of magnitude (e.g., their Figure 1), 
or decrease by more than an order of magnitude (e.g., their Figure 3) 
in the inner few AU region. 
We translated this into an enhancement or decrement in the 
oxygen abundance
by up to an order of magnitude relative to the standard case. 
We describe the results of our model calculations for
the reference case in Section 3.1 and the results for the
variations in Section 3.2.

\section{Results}

\subsection{Reference Case} 

Figure 1 shows the vertical variation of some of the basic physical
quantities of the disk atmosphere  for a range of radial distances
in the inner disk between 0.25\,AU and 20\,AU.  In all three panels, the 
abscissa is the vertical column density $\NH$, integrated downward
into the disk from above. The top panel shows the ionization rate
multiplied by the inverse-square dilution factor.  
The similarity of the profiles at the different radii shown 
illustrates the usefulness of 
characterizing disk properties as a function of vertical column. 
Note that 
Compton scattering, which is not treated here, 
begins to become important for column densities approaching 
the maximum value in the figure $\NH = 10^{24} \persqcm$. 

The temperature variation with column density (Fig.~1; middle
panel) shows a particularly rapid drop at vertical column densities
$\NH \sim 0.6-1.5 \times 10^{21} \persqcm$. The temperature is high,
in the range 4000--5000\,K, at column densities below this value, but it
then drops into the warm 200--2000\,K range at larger column
densities.  The atmosphere remains warm until collisions with dust
grains bring the gas temperature down to the dust temperature; this
occurs somewhere in the column density range 
$\NH = 10^{22} - 10^{23}\persqcm$, with the value depending on radius.  
The 200--2000\,K region (henceforth the ``warm atmosphere'') is a 
region of high density with $\nH =3\times 10^9 - 3\times 10^{10}\,\percc$ 
in the reference case.
As shown in the bottom panel of Fig.~1, the ionization fraction 
decreases by factors of $\sim 100$ where the gas temperature drops 
abruptly and the transition from H to $\Htwo$ occurs. 

Many molecular abundances also increase by large amounts as the temperature 
and ionization fraction drop, led by the atomic to molecular hydrogen 
transition shown in the bottom panel of Figure 1. The change from a primarily 
atomic to a primarily molecular medium, and from higher to lower levels of 
temperature and ionization, are the manifestations of a {\it thermal-chemical}
transition.  This transition is mediated thermally by the onset of strong 
molecular line cooling that causes the sharp decrease in temperature. 
The chemical transition to molecules results from the decrease in the ionization 
level of $\Hp$ and $\Hep$, which destroy molecules, and the suppression of 
chemical temperature-dependent reactions that destroy them. 

Thermal-chemical transitions are commonly found among models in the literature, 
although the details of the transition may differ depending on the 
irradiation field and other model properties. The sharpness of the 
transition is one property that can 
vary depending on the model parameters, 
and sharp transitions like that of the (X-ray irradiated) reference model 
may appear in UV irradiated models as well.
The UV irradiated model of Woitke et al.\ (2009) is a good example.
The electron fraction in that model varies sharply, from $10^{-4}$ to
$10^{-5}$ at $z/r = 0.15$ over a radial range from $\sim 0.5$ to 10\,AU.  
The specific details differ from our model, although the possibility of 
sharp transitions is a common element.
Models that include both UV and X-rays also find results similar
to ours, e.g., a recent calculation by Woitke (2011, personal communication) 
has a similar temperature structure.  Nomura et al.\ (2007) also 
find a sharp temperature decline when X-rays and UV are considered and 
grains have grown to large size.
There is also diversity among X-ray irradiated models: we find sharper 
transitions when mechanical heating is included, and shallower transitions 
when mechanical heating is weak (Glassgold et al.\ 2004, 2009; see also 
the following section).  

The top two panels 
of Figure 2 show the transitions of carbon, oxygen and nitrogen to 
molecules for the reference model at radii 
0.25, 1.0, and 2.0\,AU (Figure 2a), and 4.0, 10, and 20\,AU (Figure 2b). 
The abundances in this figure are defined relative to that of all hydrogen nuclei,
and the region below the main thermal-chemical transition near 
$\NH \sim 1-3 \times 10^{21} \persqcm$ is emphasized.  
Focusing for the moment on the results for $r = 1$\,AU, 
we find that, like the transition from atomic H to $\Htwo$, 
the transition from atomic carbon to CO is complete,  
and CO remains abundant down to the largest vertical columns. In contrast, 
the conversion of all of the residual oxygen (that is not in CO) into $\HtwoO$ 
occurs in the warm atmosphere ($\NH = 1-3\times 10^{21}\persqcm$), but the 
$\HtwoO$ abundance declines at larger columns where the gas is no longer warm, 
with $\Otwo$ and $\SOtwo$ as well as atomic O becoming the major reservoirs of 
the residual oxygen.  

Several other recent thermal-chemical calculations also study $\HtwoO$ 
synthesis in disk atmospheres.  In the model of Woitke et al.\ (2009), 
water is present in the disk atmosphere at 1\,AU although at 
much lower column densities ($< 10^{14}\,\persqcm$ above 200\,K) than 
found here.  In the calculation of Willacy \& Woods (2009), a much larger 
column of warm water ($\sim 10^{19}\,\persqcm$) is present and at higher 
temperature ($\gtrsim 1000$\,K). 

As shown in the second panel (from the top) of 
Figure 2, the transformation from 
atomic to molecular nitrogen becomes 
complete in the warm zone, and atomic N remains fairly abundant throughout much 
of the inner disk. The bottom panel shows the molecular ions generated by X-ray ionization. Although they are important for the chemistry, they do not dominate 
the ionization balance, which is accomplished in this model by heavy atomic ions 
such as ${\rm Si}^+$, ${\rm Fe}^+$ and ${\rm Mg}^+$. However, this conclusion
is sensitive to the assumed elemental abundances, as discussed by AGM11.

We find that a thermal-chemical transition 
from hot and atomic to warm and molecular 
always occurs in X-ray irradiated disks, at least out to 40\,AU,
although the details of the transition 
(such its location, sharpness, and other properties) 
are sensitive to the input parameters of 
the chemical and thermal model. 
We discuss this topic further in Section~3.2, 
where we describe variations about the reference  model. 
In this section we focus on the molecular abundances 
found in the reference case. 

The physical and chemical properties of the disk atmosphere 
that are reported here are consistent with those from our previous models 
(e.g., GNI04, GMN09). 
The present work does have some quantitative differences with our earlier work.
These are associated with the many improvements in the physical-chemical 
model and the computational program employed here.

It is no accident that the transition to from atomic C to CO follows closely the
temperature and ionization transitions in Figure 1, since CO is a
dominant coolant in the thermal-chemical transition. The attainment
of maximal CO abundance occurs relatively high in the disk atmosphere
through the action of warm neutral radical reactions, as discussed
in GNI04 (Sec.~3) and GMN09 (Sec.~2.2). The formation of the OH
radical is a key step in the synthesis of CO and $\Htwo$, through 
the endothermic reaction,
\be
\label{slowestneutral}
{\rm O} + \Htwo \rightarrow {\rm H} + {\rm OH};
\ee 
the rate coefficient recommended by Baulch et al.~(1992) is
\be
\label{neutreacOH}
k = 8.5\times 10^{-20}\,T^{2.67}\, e^{-3163/T} {\rm cm}^3 \pers.
\ee
Water is then formed by an exothermic reaction with a modest barrier,
\be
\label{ohneutral}
{\rm OH} + {\rm H}_2 \rightarrow \HtwoO + {\rm H}; 
\ee 
its rate coefficient is
\be
\label{neutreacH2O}
k = 8.5\times 1.70 \times 10^{-16}\,T^{1.60}\, e^{-1660/T} {\rm cm}^3 \pers.
\ee 
By contrast, CO is formed from OH by a fast neutral reaction without
a significant barrier.  Despite the barriers in Eqs.~\ref{neutreacOH}
and \ref{neutreacH2O}, water and CO are formed efficiently by the
above reactions at temperatures as low as $\sim 250$\,K. 

This
situation is not duplicated in the case of hydride formation for
the two other abundant and chemically interesting heavy atoms, C
and N. In reactions analogous to Eq.~\ref{neutreacOH}, with O
replaced by C and N, the thermal barriers are now $\sim 12000$\,K
instead of $\sim 3000$\,K. These barriers suppress the formation
of CH and NH by reactions with $\Htwo$, and the inverse reactions,
in which atomic H destroys CH and NH, are
both fast.  In our X-ray irradiated models, the abundance of H
decreases slowly below the 
thermal-chemical transition, so both suppressed
formation 
(via reactions with $\Htwo$) 
and rapid destruction (by H; 
i.e., reactions analogous to those in Eqs.~1 and 3)
imply
that neutral production of carbon and nitrogen hydrides along the lines of
oxygen-hydride synthesis is inefficient. 

Nonetheless, the warm neutral chemistry that is responsible for most carbon 
and oxygen going into CO and $\HtwoO$ plays an important role in the synthesis 
of other molecules. This is shown in Figure 3, which is a highly schematic chemical network that shows how fast neutral reactions of OH with N, O and CO 
generate other molecules such as $\COtwo$, $\Ntwo$, and $\SOtwo$ (not shown)
and more complex molecules that will be discussed below. Thus the OH radical
is the centerpiece of the warm chemistry of protoplanetary disk atmospheres.

As shown in Figure 3, the production of NO by the fast reaction, 
\be \label{neutreacNO}
{\rm OH} + {\rm N} \rightarrow {\rm NO} + {\rm H} \hspace{0.25in} k
\sim  5\times 10^{-11} {\rm cm}^3 \pers, \ee leads to the formation
of $\Ntwo$ \be \label{neutreacN2} {\rm NO} + {\rm N} \rightarrow
\Ntwo + {\rm O} \hspace{0.25in} k \sim  4\times 10^{-11} {\rm cm}^3
\pers.  \ee 
The nitrogen hydrides can be synthesized by successive hydrogenation of 
the N$^+$ ion, produced by X-ray ionization, via successive
reactions with $\Htwo$, followed by  dissociative recombination of
the hydride ions. Ion-hydrogenation becomes weak beyond $\NHthree^+$,
but proton transfer from molecular ions to neutral hydrides will
produce hydride ions up to NH$_4^+$. Thus a combination of ion-molecule
and neutral reactions initiated by X-ray ionization can synthesize
ammonia. 
A related sequence of neutral reactions starting with NO
leads to the synthesis of HCN, 
\be \label{neutreacCN} {\rm NO} +
{\rm C} \rightarrow {\rm CN} + {\rm O} \hspace{0.25in} k \sim  3\times
10^{-11} {\rm cm}^3 \pers, \ee
\be
\label{neutreacHCN}
{\rm CN} + \Htwo \rightarrow {\rm HCN} + {\rm H} \hspace{0.05in}
k(300\,{\rm K}) \sim  2.0\times 10^{-14} {\rm cm}^3 \pers.
\ee
Although many reactions have been omitted from this brief description 
of the nitrogen chemistry for purposes of clarity,
the middle panel of Fig.~2 shows that the main C $\rightarrow$ CO
and O $\rightarrow \HtwoO$ transitions are accompanied by the almost
full conversion of N to $\Ntwo$.  The transformation from N to
$\Ntwo$ is mediated by neutral reactions as indicated in Fig.~3.
It is accompanied by significant warm columns of other nitrogen-bearing 
molecules, such as NO, $\NHthree$, HCN and N$_2$H$^+$.  

The integrated treatment of both neutral and ion-molecule reactions 
is a feature of the underlying chemical model, with some 
molecular species demonstrating a greater sensitivity to X-ray 
irradiation than other species.  
The X-rays play a critical role 
in the synthesis of $\NHthree$ and N$_2$H$^+$. 
Since FUV radiation is unable to ionize  N and $\Ntwo$ 
(which have ionization potentials of 14.5 and 15.6\,eV, respectively),
the X-rays are important in activating the nitrogen chemistry. 

The hydrocarbon chemistry proceeds similarly to the nitrogen chemistry. 
The number of hydrogen atoms in a 
hydrocarbon molecule or ion can be increased by reactions
with $\Htwo$ or decreased by reactions with H.  Neutral hydrocarbons
can be ionized by proton transfer from abundant molecular ions,
such as H$_3^+$, HCO$^+$ and H$_3$O$^+$, and neutralized by dissociative
recombination with electrons; their charge is also affected by
charge transfer processes. These types of reactions are believed
to be operative for the hydrocarbon families of interest, those with 
one or two carbon atoms.  A key point is that double-carbon molecules
are synthesized from those with just one carbon atom by {\it insertion
reactions} with C and C$^+$, as illustrated by the simplest examples,

\be
\label{insertion1}
{\rm  C} + {\rm CH}_n 	\rightarrow {\rm C}_2{\rm H}_{n-1} + {\rm H} \hspace{0.25in} (n=1-4),
\ee 

\be
\label{insertion2}
{\rm  C}^+ + {\rm CH}_n 	\rightarrow {\rm C}_2{\rm H}_{n-1}^+ + {\rm H} \hspace{0.25in} (n=1-4),
\ee
or by molecular ions as suggested by Woodall et al. (2007) 
\be
\label{insertion3}
{\rm  CH}_3^+ + {\rm C} 	\rightarrow {\rm C}_2{\rm H}^+ + \Htwo
\hspace{0.15in} {\rm and} \hspace{0.15in}
{\rm  CH}_2^+ + {\rm C} 	\rightarrow {\rm C}_2{\rm H}^+ + {\rm H}.  
\ee

These highly simplified remarks about hydrocarbon chemistry are
closely aligned with the general discussion of Ag\'undez, Cernicharo
\& Goicoechea (2008).  In their model variation that is closest to our 
model, these authors
use a photodissociation code to calculate the chemical structure
of the surface region of the D'Alessio et al.~(1999) disk model. They assume
that the gas and dust temperatures are the same and that the dust
is interstellar, and they consider stellar as well as interstellar UV
radiation but ignore X-rays. 
Despite the many differences with their
model, we draw the same general conclusion: hydrocarbons are predicted
to be present in the top layers of protoplanetary disk atmospheres.
There are of course quantitative differences between the results
of Ag\'undez et al.\ and those reported here. We emphasize warm
molecular column densities that arise in regions with $T \geq
200-300$\,K, because they are related to measured fluxes.  Ag\'undez
et al.\ do not calculate gas temperatures, and they report total
vertical column densities in the disk surface layer, almost all of 
which is cool gas ($<300$\,K) at 1\,AU.

Willacy \& Woods (2009) also synthesize a signficant column of warm 
HCN.  At 1\,AU they find a column of $\sim 4\times 10^{16}\,\persqcm$ 
at a temperature $\gtrsim 1000$\,K.  
As they note, the much lower columns of $\CtwoHtwo$ that result in 
their models ($\sim 9\times 10^8\,\persqcm$) are likely due to 
the omission of reactions involving $\Htwo$ with C$_2$ or 
C$_2$H.

Despite the fact that CO tends to take up most of the available carbon, 
the hydrocarbon chemistry is still active after the main transition.
This is a characteristic feature of the X-ray chemistry, where X-ray 
generated He$^+$ ions produce C$^+$ ions when they destroy CO by 
the ion-molecule reaction, 
\be 
\label{He+plusCO} {\rm  He}^+ + {\rm CO}  \rightarrow
{\rm C}^+ + {\rm O} +  {\rm He}. 
\ee 
Similar reactions, in which
He$^+$ generates atomic ions from other molecules, are in fact quite
common, notably for N$_2$ and CO$_2$. In the latter case O$^+$ is also
produced.

In general, we find that both neutral and ion-molecule reactions
contribute to the production and destruction of molecular species,
with the relative importance of the two kinds of reactions varying
with position in the disk as well as with species.  Neutral reactions
are more important in the warm disk surface, while ion-molecule
reactions play a significant role in the cooler regions below. The 
species shown in Fig.~2 offer many challenges 
for the detection of molecules in the inner parts of protoplanetary 
disks, and some have already been observed, as discussed in Sec.~1. 

We briefly summarize the results for the reference model at 1\, AU,
describing our results for the more abundant molecules.  To do this, 
we take advantage of the monotonically decreasing run of temperature
with increasing vertical column density (Fig.~1) and integrate the
species' volumetric densities downward from the maximum (2000\,K) 
to minimum (200\,K) temperature of the warm atmosphere.  
This corresponds to the
region just below the main thermal-chemical transition discussed
in Section~2 and roughly to the region that can produce observable
line emission in disks seen partly face-on.  We often refer henceforth 
to these integrated columns simply as ``warm columns''.
                                 
\noindent{$\bullet$} $\HtwoO$ forms in the same region as CO and reaches a comparable abundance, $\sim 10^{-4}$.  The water abundance declines below 
$4\times 10^{21}\persqcm$ before recovering its high abundance approaching  
$10^{24}\persqcm$.  The resulting column density of warm water is 
$\sim 5\times 10^{17}\persqcm$.  In the region of reduced water abundances, 
several other species take up the available oxygen.  
Together with water they (i.e., $\Otwo$ and  SO$_2$) would enhance the 
cooling above that provided by O and  CO, the two species that are 
contribute to the cooling in the present calculation.  \\
\noindent{$\bullet$} HCN is
less abundant, reaching a maximum abundance of $\sim 10^{-7}$  in
the warm atmosphere.  The total column of warm ($> 200$\,K) HCN is
$\sim 10^{15}\persqcm$. \\ 
\noindent{$\bullet$} $\CtwoHtwo$ is less
abundant than HCN, reaching a maximum abundance of  $10^{-8}$ in
the warm atmosphere. The column of warm $\CtwoHtwo$ is $\sim
10^{13}\persqcm$.    \\ 
\noindent{$\bullet$} $\COtwo$ has an abundance
in the range  $10^{-6}-10^{-8}$ in the warm atmosphere; its warm
column density is $\sim 5\times 10^{15}\persqcm$.   \\ 
\noindent{$\bullet$}
HCO is more abundant than $\COtwo$, with a warm column density of
$\sim 1.4\times 10^{16}\persqcm$. The warm column density of HCO$^+$
is $\sim 1.4\times 10^{13}\persqcm$  \\ 
\noindent{$\bullet$}
$\NHthree$ reaches its maximum abundance of $\sim 10^{-6}$ just
below the warm region; its warm column is  $\sim 3\times 10^{15}\persqcm$

While $\HtwoO$ and some other molecules such as HCN and $\CtwoHtwo$
form high up in the atmosphere and then decline in abundance, some
species increase further down (e.g., $\CHfour$), or at least do not
decrease ($\NHthree$; Fig.~2). $\CHfour$ has a maximum abundance in the
middle of the warm region where $T \sim 800$\,K. Just below the
warm region, it begins to recover from its abundance low 
($<10^{-10}$) and becomes abundant at large depths: $x(\CHfour)
\sim 10^{-7}$ for  $ \NH \gtrsim 10^{23}\persqcm$. We find that the 
chemical as well as the temperature and ionization structure of our disk
model is highly stratified. 
Such a stratified structure is unlikely to be altered by turbulent 
mixing unless the transport timescale is unusually short 
(e.g., Heinzeller et al. 2011).

\subsection{Variations on the Reference Case}

In the previous section we presented the abundances of major species
obtained for the reference case defined in Table 2.
Even for fixed stellar and disk properties, the model is characterized 
by several parameters: the
mechanical (accretion) heating, the dust surface area, the C/O ratio
and the radial distance. Here we investigate how the variation of 
these parameters affects the predictions of the model.

One important uncertainty is the efficiency of $\Htwo$ formation
on grains, as discussed at the beginning of Section~2.  Without it, i.e.,
with only gas phase formation of $\Htwo$, the abundances of molecules
other than CO ($\HtwoO$, $\COtwo$, HCN, $\CtwoHtwo$) are much reduced
in the warm atmosphere, with the result that their warm columns are
very small (see also Ag\'undez et al.\ 2008; GNI04).  
As noted in Section~2, theoretical calculations do support $\Htwo$ 
formation on {\it warm} grains, although there is still considerable 
uncertainty about the efficiency of this process (e.g., Cuppen et al.~2010).
The formation rate also depends on the surface area of the
grains, which we parameterize in terms of $a_g$ (defined in Sec 2). 
The grain surface area in the reference model is smaller than 
the ISM value by a factor of $\sim 30$. 

Figure 4 compares the thermal-chemical transition of the reference case 
(black curves with $a_g = 0.707 \mu$m, $\alpha_h = 1.0$) and two other cases
in which $a_g$ is increased by a factor of 10 (red curves) and $\alpha_h$ is 
reduced by 100 (blue curves). The red curves correspond to a reduction in the 
grain surface area per hydrogen nucleus by a factor of $\sim 300$ 
relative to interstellar grains, and the blue curves have essentially 
no mechanical heating. In the former 
case, the transition is shifted significantly deeper down into the disk, from 
$1 \times 10^{21} \psqcm$ to $5 \times 10^{21} \psqcm$, with a corresponding 
increase in volumetric density. Because the dust cooling rate is also reduced
by a factor $\sim 300$, the depth of the warm layer is extended, and the 
warm molecular columns, given in Table 3, 
are all increased, with the exception of
OH. When mechanical heating is essentially eliminated, the transition is smoother 
and shifted upwards in the atmosphere to the region $\sim 2 \times 10^{20}\persqcm 
- 5 \times 10^{20} \persqcm$. The associated warm region is also shifted upwards, 
and the warm molecular columns are decreased by a factor of five to ten, 
with the exception of OH, which is increased by a factor of about three.

From this discussion of we can conclude that warm molecular columns increase 
with increasing dust size (i.e., reduced dust surface area; parameter $a_g$) 
and increasing mechanical heating (parameter $\alpha_h$). Both effects 
are accompanied by a downward shift in the location of the atomic to molecular 
transition.  The warm region starts at the transition, and larger warm columns  
(which favor detection) are 
generated as the transition moves downward into regions of higher density.  
Table 3 gives the warm columns for 
some of the more abundant species that have either have already been observed 
in T Tauri disks or might be candidates for future detection.

X-rays are the only source of external radiation in these model calculations.  
While varying $\Tx$ from 1\,keV to 5\,keV did not much alter our results, 
variations in $\Lx$ have a much larger effect.  
Table 4 shows how the warm column of selected molecules and ions 
depends on $\Lx$ at 1\,AU for a large range of values of $\Lx$ when the 
other model parameters are kept the same as in the reference case. 
From {\it Chandra} observations of young clusters, Preibisch \& Feigelson (2005) found that the X-ray luminosity function depends mildly on cluster age and, within clusters, on the stellar mass range. For the Orion Nebular Cluster, the median K star X-ray luminosity is $\Lx = 3\times 10^{30}\,\ergpers$, with 90\% of the stars having
$\Lx > 4\times 10^{29}\,\ergpers$ and 10\% having $\Lx = 1.3\times 10^{31}\,\ergpers$. 
In Table 4 we can see that
the warm columns of ions are especially sensitive to the X-ray
luminosity and, as might be expected, increase dramatically as $\Lx$
increases. 
Even for X-ray luminosities as low as $\Lx = 10^{26}\,\ergpers$ 
(i.e., several orders of magnitude lower than the X-ray luminosities of 
typical T Tauri stars), the X-rays are a significant source
of ionization, and could influence the chemistry of T Tauri disks
by producing a variety of molecular ions by proton transfer and by
destroying neutral molecules by reactions with H$^+$ and He$^+$.
At these low X-ray ionization rates, the stellar UV radiation field 
(not treated here) would also be particularly important.

But simple neutral molecules change little as $\Lx$ goes
from $10^{26}$ to $10^{31}\,\ergpers$. The reason is that mechanical
heating associated with the MRI theory of viscosity dominates the
heating of the disk atmosphere, and molecular synthesis in the warm
regions is built on temperature-sensitive neutral reactions, as
discussed in the previous section. 
In the presence of strong accretion 
heating, the X-rays also have little effect on the location of the main 
thermal-chemical transition defined by CO and $\Htwo$.  
For example at $R=1$\,AU, the transition shifts 
from $6.5 \times 10^{20}\persqcm$ for $\Lx = 10^{28}\,\ergpers$ to 
$1.0 \times 10^{21}\persqcm$ for $\Lx = 10^{31}\,\ergpers$, while
for $\Lx < 10^{28}\,\ergpers$ there is no change at all.
Nonetheless, molecular ions still play a role,
sometimes in forming neutrals by dissociative recombination and
sometimes in destroying them by proton transfer or charge transfer
reactions.

By way of contrast, radicals are sensitive to X-rays, as exemplified
by OH and NO, both of which increase by about 3,000 when $\Lx$ is
increased from $10^{26}\,\ergpers$ to $10^{31}\,\ergpers$. 
In our model, the steady state abundance of OH
in warm regions is mainly determined by its production by recombination
of H$_3$O$^+$ and destruction by neutral reactions. NO follows OH,
since it is produced from OH via Eq.~\ref{neutreacNO} and destroyed
by atomic N via Eq.~\ref{neutreacN2}. A similar explanation applies
to SO and SO$_2$, which are moderately sensitive to X-rays. It is
of considerable interest that the warm columns of the hydrocarbons
$\CHfour$ and $\CtwoHtwo$ 
decrease with increasing $\Lx$. This trend is stronger for $\CHfour$,
very likely due to the fact that it is produced by dissociative
recombination of the molecular ion CH$_5^+$ which is produced by
the radiative association of CH$_3^+$ an $\Htwo$.

Variations in the O/C ratio can have a dramatic effect on the abundances
in the warm disk atmosphere. Table 5 gives examples for the
reference case at 1\,AU. The oxygen abundance has been varied at 
fixed carbon abundance so that the O/C ratio varies from 0.25
to 5.0; the reference case value is 2.5. Below O/C = 1, the reduction
in the available oxygen leads to a significant decrease in the warm 
column of CO and also of $\COtwo$.  Pure oxygen species such as 
O, OH, $\HtwoO$, $\Otwo$, H$_3$O$^+$ show very large decrements, 
especially below
O/C = 1, as do closely related oxides such as $\SOtwo$ and NO. At the
same time, pure carbon species like $\CtwoHtwo$ and $\CHfour$
increase drastically. Nitrogen species also change, but by large
amounts only when they contain carbon (e.g., HCN) or oxygen (e.g.,
NO). One noteworthy result of this analysis is that the warm column
of Ne$^+$ increases rapidly with decreasing O/C ratio because it is
efficiently destroyed by $\HtwoO$.

Ratios of warm columns derived from molecular line flux measurements 
are often used in making preliminary interpretations of the observations. 
Our results show that variations in the O/C ratio can cause significant 
changes in the warm column ratios. Because $\HtwoO$ decreases more rapidly  
than CO with decreasing O/C, the $\HtwoO$/CO ratio becomes very small 
for small O/C  $\leq 1$. By way of contrast, the HCN/$\HtwoO$ ratio 
is a sensitive decreasing function of the O/C ratio and becomes large 
as O/C approaches unity. The same is true for the $\CtwoHtwo$/$\HtwoO$ ratio, 
but even more so because the warm column of $\CtwoHtwo$ is a much more  
sensitive function of O/C than HCN. Thus these ratios can be considered 
as signatures of enhanced carbon chemistry.

Looking at how the overall abundance patterns vary with disk radius, we
find that the abundance patterns at 
1\,AU are similar at smaller and
larger radii, as demonstrated in Figure 2 for oxygen molecules 
(top panel), nitrogen molecules (middle panel) and molecular ions (bottom
panel). Some species show large variations in their vertical column density 
distributions as a function of 
radius, e.g., $\COtwo$.  
Our results for 40\,AU are not plotted because they are qualitatively 
similar to those for 20\,AU.
Figure 5 and Table 6 show the radial variation of 
warm columns (i.e., the regions of the disk where line 
emission can be efficiently generated) for the reference case. 
The radial range in Figure 5 and Table 6 extends only to 10\,AU 
because the disk atmosphere at 20--40\,AU does not 
have significant column density in the 200-2000\.K 
range (Fig.~1). 
Note that the gas column densities shown in Figures 2 and 5 and 
Table 6 do not include the effect of freeze-out, which would 
have a greater impact in the disk midplane.

We see that the warm columns are large at very small radii,
decrease rapidly inside 1\,AU, and then decrease more slowly
beyond. 
This behavior is typified by $\Htwo$ itself and is shared by many of 
the more abundant molecules,
e.g., CO, $\HtwoO$, $\Otwo$, $\NHthree,$ and HCN.
The water column increases significantly at radii $< 0.5$\,AU because
the neutral reactions that produce water are effective throughout
the warm atmosphere (because of the high gas temperature and density
in this region) and lead to a high abundance of water throughout
(Fig.~2).  In contrast, at 1\,AU, the water
abundance drops at intermediate heights 
(below $\NH=5\times 10^{21}\persqcm$) before recovering deeper down, 
leading to a lower warm column than at 0.25\,AU (Fig.~5).

It is interesting that, in the top panel of Figure 5, the large warm 
column of atomic H changes by no more than a factor of two from 
0.25\,AU to 10\,AU. The reason 
for this is that the basic OH chemistry operative in the warm region  
(Eqs.~\ref{slowestneutral} and \ref{ohneutral}) releases atomic H as 
atomic oxygen is processed into water. The abundance of atomic H is 
important for the synthesis of hydrocarbon molecules because many of 
the reactions that hydrogenate carbon are reversed by atomic H reactions.

Many of the heavier molecules show similar patterns because 
they share a common (warm neutral radical) chemistry and because 
the  physical properties of the warm region change slowly with disk radius.
For example, 
the ionization parameter has the value $\zeta / \nH \sim 3 \times
10^{-20} \ccps$ to within a factor of two over radii from  
0.25 to 20\,AU (a factor of almost 100) because the inverse square
dilution of the X-ray ionization rate (top panel of Fig.~1) is
largely compensated for by the decrease of the density with radius.
Consequently the electron fraction at the top of the warm region
is almost independent of radius, with a value close to $x_{\rm e}=2\times
10^{-6}$. 
This approximate invariance applies to the ionization of the warm
region, but not to the molecular abundances which are affected by
density and temperature (and processes such as freeze out) as well
as ionization.  
The temperature
does remain close to 1000\,K at the top of the warm column for
small radii (middle panel of Fig.~1), but it then decreases
significantly beyond $4$\,AU. Of course the density at a given vertical 
column density decreases by
more than 1,000 as the radius increases from 0.25 to 20\,AU. These changes
also cause the thermal chemical transition to move slightly upward
in the atmosphere with increasing radius, and this also tends to lead to
a reduction in warm column densities.
The decreasing density of the warm atmosphere as a function of 
radius may play a role in limiting the radial extent of emission 
from molecular transitions with high critical densities.

\section{Discussion}

\subsection{Inner Planet Formation Region}

Our model results (Section 3) have potential implications for the 
understanding of the inner and outer planet 
formation regions of the disk, i.e., the regions roughly within 
and beyond the ``snow line'' where water vapor would condense 
on grains.  The inner planet formation region, discussed in this 
section, is probed by the molecular observations made with {\it Spitzer} 
and ground-based near-infrared spectroscopy. 
The outer planet formation region is discussed in the next section. 

\subsubsection{Warm Molecular Columns}

As we found in GMN09, 
the warm atmosphere in our models achieves temperatures comparable 
to that of the {\it Spitzer} molecular emission from T Tauri disks 
($\sim$300--1000\,K; Carr \& Najita 2008, 2011; Salyk et al.\ 2008, 2011).  
These temperatures are maintained over significant column densities at 
disk radii out to $\sim 4$\,AU (Figure 1, middle panel). 
Thus the temperature structure of our disk atmosphere 
is consistent with the interpretation that the {\it Spitzer} molecular 
emission arises from the inner few AU of the disk. 

Some molecular species appear quite sensitive to the parameter 
variations we considered, with their warm molecular columns varying 
significantly over the parameter range we explored.  As one example, 
the variation in grain size and mechanical heating alone can 
produce warm column variations of $\sim 30$ in $\HtwoO$ and HCN 
(Table 3).  This sensitivity may help to account for the factor of 
$\sim 30$ range in the $\HtwoO$ and HCN emission strengths among 
T Tauri disks 
(Pontoppidan et al.\ 2010; Salyk et al.\ 2011; Carr \& Najita 2011). 
The role of grain settling in producing enhanced warm columns may 
also help to account for the finding by Salyk et al.\ (2011) of 
the enhanced detectability of molecular emission from more settled disks.  

The warm molecular columns we find for some species are also quantitatively 
similar to those inferred from the {\it Spitzer} data. 
The line-of-sight column density of the water emission detected with 
{\it Spitzer} ($\sim 10^{18}\persqcm$), which has an 
associated projected emitting area of 
$\pi r^2$ where $r \sim 1$\,AU (Carr \& Najita 2011; 
see also Salyk et al.\ 2011), 
is similar to our (vertical) warm column of water at 1\,AU 
($10^{17}-5\times 10^{18}\persqcm$; Tables 3, 4, and 5).  
The much smaller warm columns achieved at larger disk radii 
($< 10^{17}\persqcm$ at 10\,AU in the reference case)   
help to account for the inference of AU-sized water emission regions 
from the {\it Spitzer} spectra, although perhaps not completely 
(see Section 4.2). 

Larger warm columns are achieved at small disk radii 
($\sim 10^{20}\persqcm$ at 0.25\,AU in the reference case; 
Table 6 and Figure 5). 
These are
reminiscent of the water columns reported at such 
radii based on $K$- and $L$-band spectroscopy of disks 
($\sim 10^{20}-10^{21}\,\persqcm$; 
Carr et al.\ 2004; Najita et al.\ 2009; Doppmann et al.\ 2011). 
However, the temperature of the observed emission is warmer 
($\sim 1500$\,K) than most of the warm column (2000--200\,K) in the model. 
Thus, while the empirical situation is that strong near-infrared water
emission is rare (e.g., Najita et al.\ 2007), this does not necessarily
mean that water is underabundant at small disk radii in most disks.  
Our models suggest that large columns of water are commonly present at small
disk radii, but the water is typically not warm enough to produce
strong water emission in the near-infrared.

We can also compare our results to those derived for the $4.7\micron$ 
CO fundamental emission from T Tauri stars, which 
has a characteristic temperature of $\sim$500--1000\,K, 
line-of-sight column density $\sim 10^{18}\persqcm$, and projected 
emitting area of radius $\sim 0.15$\,AU (Salyk et al.\ 2011; 
Najita et al. 2003).
Our reference model at 0.25\,AU 
produces a vertical CO column of $5\times 10^{17}\persqcm$ over the 
vertical temperature range 2000--500\,K, in reasonable agreement with 
the observed values. 
The decrease with radius of both the density of the atmosphere 
and CO column over this temperature range may contribute to 
limiting the radial extent of the CO fundamental emission from 
the disk. 

The {\it Spitzer} HCN emission detected from T Tauri stars 
has been characterized as having a typical temperature of $\sim 700$\,K 
(Carr \& Najita 2011; Salyk et al.\ 2011). 
If the emitting area is assumed to be the same as that of the water emission 
(i.e., $\sim 1$\,AU), the line-of-sight column density is typically 
$\sim 10^{15}\persqcm$ (Salyk et al.\ 2011). 
Fits made without that assumption yield smaller 
emitting areas ($0.3-0.6$\,AU) and larger HCN columns 
(a few $\times 10^{16}\persqcm$), 
although the lower columns of 
Salyk et al.\ are not ruled out (Carr \& Najita 2011).
In our models, HCN is abundant throughout the inner disk 
atmosphere, with a warm column of $10^{15}-10^{16}\persqcm$ at 1\,AU 
depending on the model parameters  
(Tables 3, 4, and 5), a range that encompasses the values inferred 
by Salyk et al.\ (2011).  The larger line-of-sight columns  
preferred by Carr \& Najita (2011) may not be too different from 
the warm (vertical) columns of our models 
if the HCN emission arises at disk radii closer to 0.5\,AU, 
where the warm column is larger, and   
when disk inclination effects are taken into account. 

The similar temperatures found for the {\it Spitzer} HCN and water emission 
(Carr \& Najita 2008, 2011; Salyk et al.\ 2008, 2011) suggest 
that the HCN emission arises from the same region of the disk atmosphere 
as the water emission. 
The smaller emitting area inferred for the HCN emission, compared 
to the water emission, in the best-fit models of 
Carr \& Najita (2008, 2011) might seem odd in that context.  The 
difference might indicate that the HCN emission 
is restricted to a smaller (possibly annular) region than the water emission, 
perhaps because of the molecular abundance distribution or 
excitation considerations.  
In our models, the smooth variation of the warm HCN column 
with radius and the slow decline beyond 1\,AU (Figure 5) 
does not suggest an annular distribution for the HCN.  However, 
the radial range of the HCN emission may also be restricted 
by the high 
densities needed to excite the observed transitions 
and the decline in density with disk radius in the warm atmosphere 
(Figure 2).

The $\CtwoHtwo$ emission detected with {\it Spitzer} 
has been characterized as having a $\CtwoHtwo$/HCN column density 
ratio of $\sim 0.1$ and a $\CtwoHtwo$ column of $10^{15}-10^{16}\persqcm,$ 
assuming that the $\CtwoHtwo$ emission has the same temperature 
and emitting area as the HCN emission (Carr \& Najita 2008, 2011).  
Salyk et al.\ (2011) instead assumed that the $\CtwoHtwo$ and 
HCN emission have the same emitting area as the water emission
but allowed the $\CtwoHtwo$ and HCN temperatures to differ; 
they inferred typical $\CtwoHtwo$/HCN column ratios $\sim 1$ and 
columns of $\sim 10^{15}\persqcm$ for $\CtwoHtwo$.
In our reference model, the $\CtwoHtwo$/HCN warm column ratio is 
$\sim 0.01$ and the $\CtwoHtwo$ column density is $\sim 10^{13}\persqcm$ 
at 1\,AU, both of which are smaller than indicated by the observations.
However, for the case $a_g=7.07$, 
the $\CtwoHtwo$/HCN warm column ratio is $\sim 0.1$ and 
the $\CtwoHtwo$ column density is $\sim 10^{15}\persqcm$ 
at 1\,AU (Table 3), more similar to the reported values.

The {\it Spitzer} $\COtwo$ emission has been fit with characteristic 
temperatures higher (Salyk et al.\ 2011) or lower than (Carr \& Najita 2011) 
that of water.  Carr \& Najita (2011) find that the $\COtwo$ column 
density is not well constrained by the observations. 
Salyk et al.\ (2011) report $\COtwo$ column densities 
$\sim 10^{15}-10^{16}\persqcm$, assuming the  $\COtwo$ emission has the 
same emitting area as the $\HtwoO$ emission.
In our models, the $\COtwo$ warm column is 
$\sim 10^{15}-10^{16}\persqcm$ at 1\,AU (Tables 3, 4, and 5), 
similar to the reported values. 

As in GMN09, our models are unable to account for the 
OH emission from inner disks. 
The observed OH columns are $3\times 10^{14}-3\times 10^{16}\persqcm$ 
(Salyk et al.\ 2011; Carr \& Najita 2008), higher than in our models 
($\sim 10^{14}-10^{15}\persqcm$; Tables 3, 4, and 5).
The larger OH columns are very likely the result of UV irradiation of 
the disk, which is not yet implemented here.  
UV irradiation can dissociate $\HtwoO$ to produce OH, 
potentially explaining the 
large OH columns that are observed (Bethell \& Bergin 2009; 
GMN09).
Hot OH emission, which may result from the photodissociation of 
$\HtwoO$ (e.g., Najita et al.\ 2010), has been reported from T Tauri disks 
(Carr \& Najita 2011), consistent with this interpretation.
Given the central role of the OH radical in the warm 
chemistry of disk atmospheres (Figure 3), an enhanced OH abundance 
(e.g., from photodissociation) may lead to enhanced abundances of 
species that are synthesized from OH.  

Our models address the question raised by Salyk et al.\ (2011) 
regarding the dominant carrier of nitrogen in the disk. Noting that 
the HCN abundance they infer does not account for all of the 
available nitrogen, Salyk et al.\ (2011)  
suggested $\NHthree$ or $\Ntwo$ as the possible carriers 
of the remaining nitrogen. 
They also pointed out how $\NHthree$ is unlikely to 
be the missing reservoir: given the upper limit on $\NHthree$ 
emission in the {\it Spitzer} spectra in the $10\micron$ region, 
they inferred that the column of warm $\NHthree$ is 
$< 10^{16}\persqcm$, and 
$\Ntwo$ was suggested to be the dominant reservoir of nitrogen. 
Consistent with this interpretation, 
our model predicts warm $\NHthree$ columns of 
$10^{15}-10^{16} \persqcm$ at 1\,AU (Tables 3, 4, and 5)  
and that most of the nitrogen is in $\Ntwo$. 

\subsubsection{Molecular Emission Trends}

Our models may also provide some context for the various 
molecular emission trends that have been reported in the literature. 
In their study of a small sample of T Tauri stars, Carr \& Najita (2011) 
found tentative evidence that 
the flux ratio of HCN/$\HtwoO$ emission increases with disk mass, 
increasing by a factor of $\sim 4$ over disk masses of 
$0.002-0.02\,\Msun$.  
They interpreted this result as a possible consequence of planetesimal 
or protoplanet formation.  
This interpretation arises from considering the transport of water 
in the solar nebula (Ciesla \& Cuzzi 2006).
Because icy bodies ranging from micron, to meter, to kilometer 
and larger size migrate at different rates, they can transport 
water ice to the inner disk at greater or lesser rates compared 
to the gas, reducing or enhancing the O/C ratio of the inner disk 
and possibly affecting the chemistry of the inner disk. 

That is, 
when grains in the outer planet formation region are small enough to 
couple well to the gas, water that is frozen on these grains accretes 
along with the gas into the inner planet formation region.  The 
water is returned to the gas phase when the ices evaporate, and 
the O/C ratio of the inner disk gas is representative of the disk as a whole. 
However, when the icy grains grow into meter-sized bodies, they 
migrate inward extremely rapidly relative to the gas.  When they 
evaporate, they hydrate the inner disk, and in depositing water 
and oxygen potentially enhance the O/C ratio there. 
In contrast, 
if icy material in the outer planet formation region grows into 
planetesimal or larger bodies, such large bodies cease to migrate 
(and may go on to form planets).  They thereby sequester water 
(and oxygen) in the outer planet formation region,   
leaving the inwardly migrating material dehydrated 
and oxygen-poor (e.g., Ciesla \& Cuzzi 2006).  This situation might 
both reduce the water abundance in the inner disk and, by reducing 
the O/C ratio, favor the 
formation of hydrocarbons in the inner disk.  

We might therefore expect that early in the disk evolution process 
(before significant grain 
growth has occured) the gaseous O/C ratio of the inner planet 
formation region will reflect the bulk O/C ratio of the disk.  At 
intermediate times, the inner disk may have a higher O/C than 
the disk as a whole.  At late times, the inner disk may have 
a lower O/C ratio than the disk as a whole.  The timescale on 
which a disk transitions from ``early'' to ``intermediate'' to 
``late'' depends on the disk mass: more massive disks would 
transition more rapidly through each phase. 

Thus, a coeval cluster of T Tauri stars with a range of disk masses 
may have inner disks with a range of gas-phase O/C ratios, and  
we might expect the inner disks of T Tauri stars to show a 
range of HCN/$\HtwoO$ flux ratios, as is observed (Carr \& Najita 2011; 
Salyk et al.\ 2011).  For 
T Tauri stars transitioning from the ``intermediate'' to ``late'' 
phase, HCN/$\HtwoO$ might increase with disk mass, possibly accounting 
for the tentative trend reported by Carr \& Najita (2011). 

Our results support this scenario.  As expected, a lower O/C 
ratio enhances the HCN abundance and decreases the $\HtwoO$ 
abundance in the inner disk.  A very modest decrease in the 
O/C ratio, from 2.5 (our reference case) to $\sim 1.8$ would 
raise the HCN/$\HtwoO$ ratio of warm columns by a factor of $\sim 4$. 
It would be interesting to measure other molecular ratios that 
might corroborate this picture.  We had earlier suggested the 
possibility of using $\CtwoHtwo$/$\HtwoO$ as a diagnostic 
of the O/C ratio (Section 3). 
Our results also suggest diagnostics of the ``intermediate'' 
phase of evolution.  For the case of O/C higher than the 
reference model (i.e., an O/C ratio of 5), oxygen-bearing molecules 
such as $\HtwoO$ and $\Otwo$ are enhanced 
and carbon-bearing molecules such as $\CtwoHtwo$ and HCN are 
reduced in abundance relative to the reference case (Table 5). 

An alternative explanation for the tentative trend of HCN/$\HtwoO$ 
emission with disk mass reported by Carr \& Najita (2011) 
may be an excitation effect, where HCN 
is better excited at the higher densities that would be present 
in a more massive disk.  
This scenario would require that the disk mass estimated from 
observations that probe large disk radii ($> 20$\,AU from 
submillimeter/millimeter observations) apply to the inner few 
AU of disks. 
These two possibilities (O/C ratio or excitation effect) could be tested 
by looking for trends with disk mass of emission from other molecules 
that are less sensitive to density. 
Detailed excitation modeling would also be useful in sorting out 
these possibilities. 

More perplexing is the report by Teske et al.\ (2010) 
of a possible trend of HCN flux with 
$\Lx$, with HCN flux increasing on average by a factor of $\sim 4$ 
in the range $\Lx= 2\times 10^{29}- 2\times 10^{30}\ergpers$.  
In our models, a higher $\Lx$ enhances the abundance of He$^+$, 
which can break up CO and potentially enhance the HCN abundance.  
However, 
our warm HCN columns vary negligibly 
with $\Lx$ in the T Tauri luminosity range 
($\Lx = 10^{28} - 10^{31}\,\ergpers$). 
It would be interesting to see if studies of larger samples 
of objects confirm a trend of HCN flux with $\Lx$. 

There is a bigger difference between the warm HCN column for the 
reference case and for very low values of $\Lx$.  
For $\Lx = 10^{26}\,\ergpers$, the warm HCN column is an order 
of magnitude larger than in the reference case (Table 4).  
As T Tauri stars are known to be strong emitters of 
X-rays at a much higher level than $10^{26}\,\ergpers$, 
the ions produced by T Tauri X-ray emission appear to 
have the effect of limiting the HCN abundance and emission from disks.  

There is tentative evidence that HCN flux increases with 
stellar accretion rate, with HCN flux increasing from 
$< 1$\,mJy-$\mu$m to $\sim 4$\,mJy-$\mu$m 
over the range $10^{-9}-10^{-7}\Msunperyr$ (Teske et al.\ 2010).  
One possible explanation for this is that the increased HCN flux 
arises from increased heating. 
GMN09 had earlier discussed how enhanced 
accretion-related mechanical heating 
in the disk atmosphere could enhance the water emission from the 
disk, primarily by increasing the column density of the warm 
atmosphere.  
We would expect a similar enhancement in HCN emission, since a 
warmer disk atmosphere would enhance the emission from other 
molecules as well as water, all other things being equal.

Our results show that indeed increased accretion-related heating 
enhances the column of warm HCN in the disk atmosphere.  The 
warm HCN column increases by a factor of $\sim 3$ 
(from $2.4\times 10^{14}\persqcm$ to $7.7\times 10^{14}\persqcm$)
for values of $\alpha_h=0.01$ to 1, with most of the 
increase in column occuring in the range $\alpha_h = 0.1-1$.  
In the higher $\alpha_h$ case, the warm HCN also occurs deeper 
in the atmosphere and at higher density where it may be better 
excited. 
If higher stellar accretion rates are accompanied by higher rates 
of mechanical heating in the disk atmosphere, 
both the increase in the warm column of HCN and the higher density 
region in which it is located may account for the increased HCN flux 
that is observed at higher stellar accretion rates. 

An alternative explanation is that the higher UV fluxes associated
with accretion lead to greater grain photoelectric heating and
enhanced warm molecular columns.  Detailed modeling is needed to
explore the impact of increased UV irradiation and whether the
higher heating and photodissociation rates lead to higher or lower
warm molecular columns.  The result is likely to depend on the
degree of grain settling in the atmosphere (e.g., Nomura et al.\ 2007).

An alternative chemical explanation for the trend is that 
the higher UV fluxes dissociate $\Ntwo$ in 
favor of atomic N, which then enhances the formation and abundance of HCN. 
Such a scenario was discussed by 
Pascucci et al.\ (2009) who argued that the higher flux ratio of 
HCN/$\CtwoHtwo$ in the spectra of T Tauri stars compared to 
those of brown dwarfs may result from the higher accretion 
rates and consequently higher UV fluxes of T Tauri stars.  
In our models, a significant fraction (approximately half) of the 
nitrogen is already in atomic form in the warm atmosphere 
even in the absence of UV irradiation, 
with the remainder primarily in $\Ntwo$. 
Including the UV dissociation of $\Ntwo$ may then have a 
limited impact on the HCN abundance, although detailed 
calculations are needed to be certain. 

\subsubsection{Water Molecule Shielding and Heating}

The column density over which UV irradiation affects the disk atmosphere,  
either chemically (through photodissociation) or 
thermally (by depositing energy into the atmosphere),  
is governed by the abundance of UV-absorbing dust and
molecules in the disk atmosphere.
Molecules may
provide the dominant source of UV opacity in T Tauri disk atmospheres
given the high inferred rate of dust settling from disk atmospheres,
as previously noted in the specific context of water by 
Bethell \& Bergin (2009). 

Our results support this interpretation. 
Since water absorbs UV photons in the 1200--1800\,\AA\ range with a cross-section
of $10^{-17} - 10^{-18}\,{\rm cm}^2$ (Parkinson \& Yoshino 2003),
stellar UV photons encounter an optical depth of unity over 
a line-of-sight disk column density (to the star) 
of $N_{\HtwoO} \sim 10^{17} - 10^{18}\persqcm$, or a vertical column
density that is $\sim 10$ times smaller. 
The much larger {\it total} vertical water column densities that can be 
synthesized in our disk atmosphere models ($\gg 10^{18}\persqcm$),
the warm column component of which ($\sim 10^{18}\,\persqcm$) is 
observed in emission from T Tauri disks
(Carr \& Najita 2008, 2011; Salyk et al.\ 2008, 2011),
appear to be capable of shielding the lower disk atmosphere and the
disk midplane from UV irradiation.

As noted above, UV irradiation of the disk may also heat the disk
surface through the absorption of UV photons by water and other
molecules. A simple estimate of the heating rate per unit volume
due to the absorption of UV photons by water is
$$\Gamma_{\rm UV} = \zeta_* \,
e^{-\tau_{\rm uv}}\,E_{\rm UV}\,x_{\rm \HtwoO}\,n_H, $$ 
where $\zeta_*$ is the unshielded dissociation rate due to stellar 
UV, $\tau_{\rm uv}$ is the attenuation of the UV along a line of 
sight to a given position in the disk, and $E_{\rm UV}$ is the energy 
deposited as heat per dissociation. Laboratory experiments show that 
water is strongly dissociated in the wavelength range from 
1200--1800\AA\  (e.g.,  Parkinson \& Yoshino 2003). 
Since the mean photon wavelength for dissociation is 
$\sim 1500$\,\AA\ and the threshold is 2340\AA, 
$E_{\rm UV} \simeq 3$\,eV. Bergin et al.~(2003) showed that the stellar 
UV flux of the representative T Tauri star BP Tau at a distance of 
100\,AU is $\sim 540$ times the standard Habing interstellar radiation 
field, or $\sim 5.4\times 10^6$ larger at 1\,AU. 
Multiplying the unattenuated interstellar photodissociation rate of 
$5.9 \times 10^{-10}\,\pers$ (Roberge et al.\ 1991) by 
$3\times 10^6$, after correcting for the use by Roberge et al.\ of 
the Draine interstellar radiation field, 
yields $\zeta_* = 1.8 \times 10^{-3} \pers$.

Similarly, the X-ray heating rate is
$$\Gamma_{\rm X} = \zeta_{\rm X} \,e^{-\tau_{\rm x}}\,E_{\rm X}\,n_H,$$ 
where $\zeta_{\rm X}$ is the unshielded X-ray ionization rate, 
$\tau_{\rm x}$ is the line of sight X-ray attenuation factor, 
and $E_{\rm X}$ is the heating per X-ray ionization. 
In our 
reference
model, $\zeta_{\rm X} = 8\times 10^{-9}\,\pers$ at the top of 
the atmosphere at 1\,AU, and $E_{\rm X} \sim 20$\,eV (GNI04). 
Substituting the given numbers and assuming that $x(\HtwoO)=10^{-4}$, 
the ratio of the UV water to X-ray heating rates is  
$\sim 3 e^{-\tau_{\rm uv}}/e^{-\tau_{\rm x}}$. 
This rough estimate confirms that UV water heating is potentially
significant at the disk surface and that it is sensitive to the
absorption of the UV by dust and by water itself.  In the reference
model of this paper at 1\,AU, accretion heating in the neighborhood
of the transition region dominates X-ray heating by a factor between
10 and 100.  A UV heating rate that is $\sim 3$ times the X-ray
heating rate could make a quantitative difference, especially where
the transition occurs.

Thus water in the disk atmosphere may heat itself 
(cf.\ Bethell \& Bergin 2009), a situation that would help 
to explain why the water emission observed with {\it Spitzer} 
has similar properties from source to source (Carr \& Najita 2011).
If water is a significant heat source at the disk surface, 
and if HCN is abundant at the disk surface 
(e.g., in the low $\Lx$ case or the case of $a_g=7.07\micron$),
UV heating by water may have a (thermal) role in explaining a
trend between HCN flux and stellar accretion rate.
In addition to its potential role as a heat source, water is also
known to be an important coolant based on its observed
emission luminosity (Pontoppidan et al.\ 2010).
It would be important to explore the impact of water
on both the heating and cooling of the disk atmosphere.
Similar considerations apply to FUV absorption by $\Htwo$, which 
is more abundant than $\HtwoO$ but absorbs over a more limited 
wavelength region.  A detailed investigation is needed to 
determine how and where these processes affect the disk properties.

\subsection{Outer Planet Formation Region} 

The outer planet formation region, with its low dust temperature 
(e.g., $<150$\,K at 10\,AU; Fig.~1), is more difficult to treat with 
our current model because 
we do not include freeze out and desorption processes. 
Moreover, the lower gas temperatures and densities in the outer 
planet formation region  
(Fig.~1) lead to slower chemical timescales.  As a result, the 
steady state abundances that we calculate may be less valid, 
and transport processes (e.g., Willacy et al.\ 2006; Willacy \& Woods 2009; 
Nomura et al.\ 2009; Heinzeller et al.\ 2011), 
which are not treated here, important.
Nevertheless, some aspects of our results may bear on existing 
observations and future attempts to observe this region of the disk. 

Our simple models predict a dramatic decrease in the warm column of water 
beyond a few AU.  At 10\,AU and beyond, much of the water that is produced 
would freeze out at the low grain temperatures in the atmosphere. 
However, the model does predict a detectable water column at smaller 
radii in the region of the disk atmosphere where the dust temperature is 
high enough that water would not freeze out.  At 4\,AU, the column of 
water above the point where the grain temperature reaches 150\,K, 125\,K, 
and 100\,K is $2\times 10^{16}\persqcm$,  
$2\times 10^{17}\persqcm$, and 
$3\times 10^{17}\persqcm$ respectively. 
Such a warm water column would likely produce detectable emission, 
in potential conflict with the {\it Spitzer} results,   
which find a small emitting area for the water emission of 
$\pi\,r^2$ 
where $r < 2$\,AU (Carr \& Najita 2011; Salyk et al.\ 2011).  
This point was made by Meijerink et al.\ (2009) based 
in part on the assumptions of the GMN09 model.

Meijerink et al.\ suggested vertical transport as a way of 
resolving the potential conflict (see also Pontoppidan et al.\ 2010). 
In the scenario they proposed, water from the disk surface is transported,  
via vertical turbulent diffusion, further down in the disk to a region 
where the dust temperature is low enough that water can freeze 
out onto the grains.  These icy grains settle to the 
disk midplane and are incorporated into large enough bodies that 
they are not relofted to the disk surface.
Meijerink et al.\ argue that this effect (a ``cold finger''), can 
account for the small emitting area inferred from the 
{\it Spitzer} water observations.  
They further suggested that the radial distribution of water emission 
at the disk surface might trace the location of the midplane 
snowline, which is of interest for planet formation. 

Since this picture, created to account for the properties of the 
water emission observed with {\it Spitzer}, would also apply to 
other molecules with similar condensation temperatures, 
one way to test the picture is to see if such molecules 
have a similar radial distribution to that of water.  
One possible test molecule is HCN; its condensation temperature 
is similar to that of water (e.g., Walsh et al.\ 2010) 
and our models predict it to remain 
abundant in the disk atmosphere out to 10\,AU or more 
in the absence of freeze out. 
Thus far, the emitting areas inferred for HCN emission from the 
{\it Spitzer} data are not sufficiently constrained to test this 
picture, as the HCN area can be smaller than, similar to, or greater 
than the water emitting area, given the uncertainties 
(Carr \& Najita 2011). 
Searches for emission from cooler water and HCN, e.g., 
with the {\it Herschel Space Telescope}, may better address this issue.

There are several considerations that would bear on whether our 
model results really motivate a ``cold finger'' for water or any 
other molecule, as differences between the model results and observations 
might be accounted for, to some extent, by some combination 
of the following considerations. 
Firstly, 
the disk atmosphere beyond a few AU may be cooler than we 
have inferred.  
If grains have settled less in the outer disk (e.g., than in the 
inner disk), we will have underestimated the gas-grain cooling
there.  
Another concern is that we have neglected water cooling.
Including such effects may reduce the warm water column and the
molecular emission from the disk atmosphere without requiring
vertical transport.
Secondly, 
the synthesis of molecules such as water may be less efficient 
than we have 
assumed.  
Some of the neutral reactions we have used have thermal barriers 
and their rate coefficients are very sensitive to temperature  
(e.g., eqs.~2 and 4) and may have been overestimated in the 
analyses of low temperature laboratory data. 
Another important issue is the (uncertain) efficiency of 
$\Htwo$ formation on grains at high temperature, as discussed 
in Section 3. 
Thirdly, 
photodissociation may limit the amount of water that is 
present in the disk atmosphere.  
While processes such as photodissociation and a reduced 
efficiency of $\Htwo$ formation on 
grains will affect both the inner and outer disks, they may 
have a greater relative impact on the outer disk (where 
the warm water column is expected to be small even without 
these effects; Fig.~5) and may push the warm water column 
below an observable level in the outer disk.

We can test the first point above by looking for molecular 
emission from molecules other than water. 
Our model predicts a thermal-chemical transition out to 
beyond 40\,AU and that CO is abundant in the disk atmosphere over 
that range of radii.  The presence or absence of any resulting 
warm CO emission (e.g., with the {\it Herschel Space Telescope}) 
would probe whether or not a warm disk atmosphere is present at radii 
beyond a few AU.

How well can the models presented here really identify 
chemical signatures of planetesimal formation, one of the longer-term 
motivations for our study?  Our variation 
of the O/C ratio in the inner planet formation region 
of the disk may capture some of the relevant behavior of the scenario
discussed in section 4.1.2. 
However, 
a detailed freeze out and desorption model and the inclusion of 
transport processes (e.g., as explored by other authors)  
is likely needed to predict which species 
remain in the disk atmosphere and at what abundance. 
There are a number of uncertainties in constructing such a 
model (see e.g., Walsh et al.\ 2010 for a discussion of 
desorption processes). 
This is an important area for future work.

One tracer of the disk atmosphere that is independent of many 
of the above concerns is HI 21\,cm radiation, which is of interest 
because one feature of X-ray irradiated disk atmospheres is 
a significant HI component.  
Our models indicate an HI column of 
$2\times 10^{17} - 7\times 10^{17} \,\persqcm$ 
in the 1--20\,AU region from each face of the disk.
A similar column extending to 100\,AU would imply an HI mass 
of $\sim 10^{28}\,{\rm gm}$, a value that is consistent with 
current HI upper limits but may be detectable with future 
observational facilities (Kamp et al.\ 2007).

\section{Summary and Future Directions}

As discussed in the introduction, one goal of our modeling effort was to 
explore the potential role of processes such as grain settling, 
X-ray irradiation, accretion-related mechanical heating, and the 
O/C ratio in determining the thermal-chemical properties of disk atmospheres.
Another was to explore how well a model that excludes UV irradiation 
can do in accounting for the existing observations of inner disk atmospheres. 
We also aimed to make detailed comparisons with the observations 
in order to determine the changes needed to improve our model.

To address these goals, we used an expanded version of our 
thermal-chemical model of disk atmospheres to explore the sensitivity 
of the properties of the atmosphere to model parameters. 
We find that the warm columns (200--2000\,K) of many molecular species are 
sensitive to grain settling and the efficiency of accretion heating, 
primarily because these parameters affect the depth of the warm molecular 
layer at the disk surface.  
We also showed how
certain model parameters (grain growth, stellar X-ray luminosity,
mechanical heating, O/C ratio) affect the molecular abundances of
disk atmospheres, and which species are sensitive to which parameters.

We find many areas of agreement with the observations.
The model parameter variations we considered can account for many of 
the properties of molecular species that have been detected
with {\it Spitzer} and ground-based near-infrared spectroscopy 
(i.e., their typical
temperatures, warm columns, and emitting areas).  The sensitivity 
of warm columns to parameters such as grain settling and accretion
heating may account, at least in part, for the large range in the molecular 
emission fluxes that have been observed from T Tauri disks.

The dependence of the warm columns on model parameters
such as grain growth, accretion heating, and O/C ratio, may help
to explain the trends reported in the literature between
the detection rate of molecular emission and mid-infrared color
(Salyk et al.\ 2011) and molecular emission flux with stellar
accretion rate (Teske et al.\ 2011) and disk mass.  Regarding the
last, we find that the ratio of HCN/$\HtwoO$ increases with a decreasing
O/C ratio; a decreasing O/C ratio in the inner disk is a potential
effect of planetesimal and protoplanet formation in the outer disk,
which likely occurs at an accelerated rate in higher mass disks
(Carr \& Najita 2011).

While the parameter variations we explored can account for the
average values and ranges of the observed molecular emission
properties of disks, a future challenge will be to explain the
properties of individual sources with a single set of model parameters.
Our model suffers from some obvious limitations; e.g., our low OH abundance 
points out the need for including UV irradiation and 
photodissociation in our model. 
The results of our steady-state chemical model 
are also likely less valid in the outer planet formation
region of the disk because of the longer chemical timescales there
and because we have not included freezeout, desorption and transport
processes.  We aim to address these challenges in future work.



\acknowledgments
We are grateful to John Carr for a careful reading of the manuscript. 
This work has been supported in part by NASA grant NNG06GF88G (Origins) and by 
NASA {\it Herschel} contracts 132594 (Theoretical Research) and 1367693 (DIGIT) to  
UC Berkeley.




\noindent \'Ad\'amkovics, M., Glassgold, A.\ E., \& Meijerink, R.\ 2011, ApJ, 
736, 143 (AGM11)

\noindent Ag\'undez, M., Cernicharo, J., \& Goicoechea, J.\ R.\ 2008, A\&A, 483, 831

\noindent Anicich, V.\ G.\ 1993, J.\ Phys.\ Chem.\ Ref.\ Data, 22, 1469

\noindent Asplund, M.,  Grevesse, N., Sauval, A.\ J.\ \& Scott, P.\ 2009, 
\araa, 47, 481	

\noindent Bai, X.-N.\ \& Goodman, J.\ 2009, \apj, 701, 737

\noindent Baulch, D.\ L.\ et al.\ 2005, J.\ Phys.\ Chem.\ Ref.\ Data, 34, 757

\noindent Bergin, E., Calvet, N., D'Alessio, P., \& Herczeg, G. J. 2003, 
\apj, 591, L159

\noindent Bethell, T., \& Bergin, E.\ 2009, Science, 326, 1675

\noindent Carr, J.\ S.\ \& Najita, J.\ R.\ 2011, arXiv:1104.0184

\noindent Carr, J.\ S.\ \& Najita, J.\ R.\ 2008, Science, 319, 1504

\noindent Carr, J.\ S., Tokunaga, A.\ T., \& Najita, J.\ 2004, ApJ, 603, 213

\noindent Carmona, A.\ 2000, Earth Moon and Planets, 106, 71

\noindent Cazaux, S.\ \& Tielens 2002, A.\ G.\ G.\ M., \apj, 575, L29

\noindent Cazaux, S.\ \& Tielens 2004, A.\ G.\ G.\ M., \apj, 604, 222

\noindent Cazaux, S., Caselli, P., Tielens, A.\ G.\ G.\ M., Le Bourlet, J.\ \&
Walmsley, M.\ 2005, J Phys, Conf. Series, 6, 155

\noindent Cazaux, S.\ \& Tielens, A.\ G.\ G.\ M.\ 2010, \apj, 715, 608 

\noindent Ciesla, F.\ J., \& Cuzzi, J.\ N.\ 2006, Icarus, 181, 178

\noindent Cuppen, H.\ M., Kristensen, L.\ E., \& Gavardi, E.\ 2010, 
MNRAS, 406, L11

\noindent D'Alessio, P., Calvet, N., Hartmann, L., Lizano, S., \& 
Cant\'o, J.\ 1999, ApJ, 527, 893

\noindent Doppmann, G.\ W., Najita, J.\ R., Carr, J.\ S., \& Graham, J.\ R.\ 2011, 
ApJ, 738, 112

\noindent Feigelson et al.\ 2005, ApJS, 160, 379

\noindent Furlan, E., et al.\ 2006, ApJS, 165, 568

\noindent Glassgold, A. E., Meijerink, R. \& Najita, J.  2009, \apj, 701, 142 (GMN09)

\noindent Glassgold, A.\ E., \& Najita, J.\ 2001, 
in ASP Conf.\ Ser.\ 244, Young Stars near Earth: Progress and Prospects, 
ed.\ R.\ Jayawardhana \& T.\ P.\ Greene (San Francisco: ASP), 251

\noindent Glassgold, A. E., Najita, J. \& Igea, J. 2004, \apj, 615, 972 (GNI04)

\noindent Gorti, U.\ \& Hollenbach, D.\ 2008, ApJ, 683, 287

\noindent Heinzeller, D., Nomura, H., Walsh, C., \& Millar, T. J.\ 2011, ApJ, 731, 115

\noindent Hirose, S.\ \& Turner, N.\ J.\ 2011, ApJ, 732, 30

\noindent Jenkins, E.\ B.\ 2009, \apj, 700, 1200

\noindent Ilgner, M., Henning, Th., Markwick, A.\ J.,\& Millar, T.\ J.\ 2004, 
A\&A, 415, 643

\noindent Kamp, I., \& Dullemond, C.\ P.\ 2004, ApJ, 615, 991

\noindent Kamp, I., Freudling, W., \& Chengalur, J.\ N.\ 2007, ApJ, 660, 469

\noindent Kress, M.\ E., Tielens, A.\ G.\ G.\ M., \& Frenklach, M.\ 2010, 
Adv.\ Sp.\ Res., 46, 44

\noindent Markwick, A.\ J., Ilgner, M., Millar, T.\ J., Henning, Th.\ 2002, 
A\&A, 385, 632

\noindent Meijerink, R., Pontoppidan, K.\ M., Blake, G.\ A., 
Poelman, D.\ R., \& Dullemond, C.\ P.\ 2009, ApJ, 704, 1471

\noindent Najita, J.\ R., Carr, J.\ S., Glassgold, A.\ E., \& 
Valenti, J.\ A. 2007, Protostars and Planets V, 
ed.\ B.\ Reipurth, D.\ Jewitt, \& K.\ Keil (Tucson: Univ. of Arizona), 507

\noindent Najita, J.\ R., Carr, J.\ S., \& Mathieu, R.\ D.\ 2003, 
ApJ, 589, 931

\noindent Najita, J.\ R., Doppmann, G.\ W., Carr, J.\ S., 
Graham, J.\ R., \& Eisner, J.\ A.\ 2009, ApJ, 691, 738

\noindent Najita, J.\ R., et al.\ 2010, ApJ, 712, 274

\noindent Nomura, H., Aikawa, Y., Tsujimoto, M., Nakagawa, Y., \& 
Millar, T.\ J.\ 2007, ApJ, 661, 334

\noindent Nomura, H., Aikawa, Y., Nakagawa, Y., \& Millar, T.\ J.\ 2009, 
AA, 495, 183

\noindent Nomura, H., \& Millar, T.\ J.\ 2005, A\&A, 438, 923

\noindent Pascucci, I., Apai, D., Luhman, K., Henning, Th., Bouwman, J., 
Meyer, M.\ R., Lahuis, F., \& Natta, A.\ 2009, ApJ, 696, 143

\noindent Parkinson, W.\ H.\ \& Yoshino, K.\ 2003, Chem.\ Phys., 294, 31

\noindent Pontoppidan, K.\ M., Salyk C., Blake, G.\ A., 
Meijerink, R., Carr, J.\ S., \& Najita, J.\ 2010, ApJ, 720, 887

\noindent Preibisch, T.\ \& Feigelson, E.\ D.\ 2005, \apjs, 160, 390

\noindent Roberge, W., Jones, D., Lepp, S., \& Dalgarno, A. 1991,
\apjs, 77, 287

\noindent Salyk, C., Pontoppidan, K.\ M., Blake, G.\ A., Najita, J.\ R., 
\& Carr, J.\ S.\ 2011, ApJ, 731, 130

\noindent Salyk, C., Pontoppidan, K.\ M., Blake, G.\ A., Lahuis, F., 
van Dishoeck, E.\ F., \& Evans, N.\ J.\ II 2008, ApJ, 676, L49

\noindent Savage, B.\ D.\ \& Sembach, K.\ R.\ 1996, \araa, 34, 279 

\noindent Semenov, D., Wiebe, D., \& Henning, Th.\ 2006, ApJ, 647, L57

\noindent Teske, J.\ K., Najita, J.\ R., Carr, J.\ S., Pascucci, I., Apai, D., 
\& Henning, T.\ 2011, arXiv:1104.0249

\noindent Wakelam, V.\ 2009, BAAS, 41, 665 [http://kida.obs.u-bordeaux1.fr]

\noindent Walsh, C., Millar, T.\ J., \& Nomura, H.\ 2010, ApJ, 722, 1607

\noindent Woitke, P., Kamp, I., \& Thi, W.-F.\ 2009, A\&A, 501, 383

\noindent Willacy, K., Langer, W., Allen, M., \& Bryden, G.\ 2006, 
ApJ, 644, 1202

\noindent Willacy, K.\ \& Woods, M.\ 2009, ApJ, 703, 479

\noindent Woodall, J., Ag\'{u}ndez, M., Markwick-Kemper, A.\ J.\ \&
Millar, T.\ J.\  2007, \aap, 466, 1197 [http://www.udfa.net]

\clearpage

\begin{center}
\begin{tabular}{|ccc|}    
\multicolumn{3}{c}{Table 1. Elemental Abundances} \\
\hline
Element 		& Abundance		& Depletion   \\        
\hline
\hline
H	& 1.00					& 1.0		\\
He	& 0.10					& 1.0		\\
C	& $1.40 \times 10^{-4}$	& 2.0		\\
N	& $6.00 \times 10^{-5}$	& 1.1		\\
O	& $3.50 \times 10^{-4}$	& 1.4		\\
Ne	& $6.90 \times 10^{-5}$	& 1.2		\\
Na	& $2.31 \times 10^{-7}$	& 7.5		\\
Mg	& $1.00 \times 10^{-6}$	& 40		\\
K	& $8.57 \times 10^{-9}$	& 12		\\
Si	& $1.68 \times 10^{-6}$	& 19		\\
S	& $1.40 \times 10^{-5}$  	& 1.0		\\
Ar	& $1.51 \times 10^{-6}$	& 1.7		\\
Fe	& $1.75 \times 10^{-7}$	& 180		\\
\hline
\end{tabular}
\end{center}

\begin{center}
\begin{tabular}{|cccc|}    
\multicolumn{4}{c}{Table 2. Model Parameters} \\
\hline
Parameter 		& Symbol	& Ref Value	 & Variation Range   \\        
\hline
\hline
Stellar Mass		& $\Mstar$	& $0.5\Msun$&                         \\
Stellar Radius		& $\Rstar$	& $2\Rsun$	 &                    \\
Stellar Temperature	& $T_*$		& 4000\,K	 &                    \\
Disk Mass		& $M_D$		&$0.005\Msun$&                        \\
Disk Accretion Rate	& $\dot M$	&$10^{-8} \Msunperyr$&                \\
Grain Size		& $a_g$		&$0.707 \mu$m&$0.707-7.07 \mu$m\\
X-ray Luminosity 	& $\Lx$		&$2\times10^{30} \ergpers$
							&$10^{26}-10^{31}\ergpers$           \\
X-ray Temperature	& $T_{\rm X}$	& 1\,keV	& 1--5\,keV           \\
Mechanical Heating	& $\alpha_h$	& 1.0	       & 0.01 --1.0            \\
O/C Abundance Ratio	& O/C 		& 2.5	  	& 0.25 -- 5          \\
Radial Distance		& $r$		& 1\,AU		&0.25--40\, AU        \\
\hline
\end{tabular}
\end{center}

\begin{center}
\begin{tabular}{|llll|}    
\multicolumn{4}{c}{Table 3. Warm Columns (200-2000\,K)$^\dagger$ } \\
\hline
$\alpha_h$   & 1.0*      & 1.0	    & 0.01   \\ 
$a_g$ 	     & 0.707*	& 7.07	    & 0.707   \\       
\hline
\hline
H$_2$        & 3.91E+21  & 2.40E+22  & 1.33E+21 \\
H$^+$        & 3.20E+12  & 3.60E+10  & 9.56E+15 \\
H$_3^+$      & 1.92E+13  & 1.45E+12  & 2.78E+13 \\
He$^+$       & 7.75E+12  & 5.00E+11  & 2.28E+14 \\
O            & 1.70E+17  & 3.70E+16  & 1.76E+17 \\
OH           & 4.57E+14  & 6.83E+13  & 1.31E+15 \\
H$_2$O       & 5.40E+17  & 5.29E+18  & 1.96E+17 \\
H$_3$O$^+$   & 4.22E+13  & 1.36E+13  & 7.50E+13 \\
O$_2$        & 3.72E+17  & 1.73E+18  & 1.05E+17 \\
CO           & 1.09E+18  & 6.55E+18  & 4.30E+17 \\
CO$_2$       & 1.26E+15  & 1.10E+16  & 2.25E+14 \\
HCO          & 1.33E+16  & 8.53E+16  & 3.54E+14 \\
HCO$^+$      & 2.28E+13  & 1.49E+12  & 4.60E+13 \\
C            & 2.68E+14  & 3.33E+13  & 9.34E+15 \\
CH$_3$       & 3.78E+09  & 2.11E+10  & 1.41E+09 \\
CH$_4$       & 9.59E+11  & 1.28E+14  & 7.35E+10 \\
C$_2$H$_2$   & 1.12E+13  & 5.53E+14  & 1.36E+12 \\
C$^+$        & 4.11E+12  & 1.33E+11  & 1.22E+15 \\
CH$_3^+$     & 5.48E+11  & 1.69E+11  & 1.41E+11 \\
C$_2$H$_3^+$ & 3.79E+08  & 3.08E+08  & 3.73E+07 \\
N            & 9.01E+16  & 1.24E+17  & 4.59E+16 \\
N$_2$        & 1.92E+17  & 1.37E+18  & 7.10E+16 \\
NH$_3$       & 3.42E+15  & 1.14E+16  & 8.40E+14 \\
HCN          & 8.05E+14  & 7.29E+15  & 2.50E+14 \\
NH$_3^+$     & 5.30E+11  & 3.50E+10  & 7.44E+11 \\
N$_2$H$^+$   & 1.53E+12  & 1.42E+11  & 2.73E+12 \\
HCNH$^+$     & 1.96E+11  & 3.67E+11  & 3.20E+11 \\
NO           & 7.29E+14  & 1.23E+14  & 1.55E+15 \\
S            & 1.28E+16  & 8.65E+16  & 1.42E+15 \\
S$^+$        & 2.42E+13  & 9.25E+11  & 6.04E+14 \\
SO           & 7.23E+14  & 3.78E+15  & 3.75E+14 \\
SO$_2$       & 9.79E+16  & 5.70E+17  & 4.21E+16 \\
CS           & 1.01E+14  & 3.87E+14  & 1.04E+13 \\
\hline
\multicolumn{4}{l}{$^\dagger$ $a_g$ in $\mu$m and warm columns in $\persqcm$.}	\\
\multicolumn{3}{l}{$^*$ Reference case.}	\\
\end{tabular}
\end{center}
\clearpage

\vspace{-0.5in}
\begin{center}
\begin{tabular}{|llllllll|}    
\multicolumn{8}{c}{Table 4. Dependence of Warm Columns (200-2000\,K) on $\Lx^\dagger$} \\
\hline
$\Lx$      &  1.0e26   &  1.0e27   &  1.0e28    & 1.0e29   &   1.0e30   &  2.0e30* &  1.0e31   \\
\hline
\hline
H$_2$      &  3.84E+21 &  3.84E+21 &  3.84E+21  & 3.79E+21 &   3.67E+21 & 3.61E+21 &  3.98E+21 \\
H$^+$      &  2.53E+07 &  3.17E+08 &  3.70E+09  & 4.37E+10 &   8.42E+11 & 2.30E+12 &  2.68E+13 \\
H$_3^+$    &  1.26E+09 &  1.45E+10 &  1.50E+11  & 1.30E+12 &   9.42E+12 & 1.62E+13 &  4.76E+13 \\
He$^+$     &  4.31E+08 &  4.83E+09 &  5.18E+10  & 4.53E+11 &   3.50E+12 & 6.39E+12 &  2.64E+13 \\
O          &  6.61E+13 &  6.55E+14 &  6.31E+15  & 4.08E+16 &   1.17E+17 & 1.40E+17 &  2.37E+17 \\
OH         &  4.03E+11 &  2.09E+12 &  1.44E+13  & 6.63E+13 &   2.57E+14 & 3.96E+14 &  1.19E+15 \\
H$_2$O     &  9.45E+17 &  8.50E+17 &  7.67E+17  & 6.43E+17 &   5.42E+17 & 5.34E+17 &  6.25E+17 \\
H$_3$O$^+$ &  3.93E+10 &  3.15E+11 &  1.65E+12  & 6.15E+12 &   2.33E+13 & 3.62E+13 &  1.22E+14 \\
O$_2$      &  2.59E+17 &  2.80E+17 &  3.17E+17  & 3.52E+17 &   3.48E+17 & 3.34E+17 &  3.44E+17 \\
CO         &  1.00E+18 &  1.05E+18 &  1.05E+18  & 1.04E+18 &   1.02E+18 & 1.01E+18 &  1.16E+18 \\
CO$_2$     &  1.77E+15 &  1.49E+15 &  1.33E+15  & 1.15E+15 &   1.13E+15 & 1.22E+15 &  1.63E+15 \\
HCO        &  9.92E+15 &  9.33E+15 &  9.25E+15  & 9.62E+15 &   1.20E+16 & 1.30E+16 &  1.15E+16 \\
HCO$^+$    &  9.30E+08 &  1.48E+10 &  2.05E+11  & 1.79E+12 &   1.15E+13 & 1.89E+13 &  5.66E+13 \\
C          &  9.56E+10 &  4.15E+11 &  2.65E+12  & 1.96E+13 &   1.34E+14 & 2.30E+14 &  7.64E+14 \\
CH$_3$     &  1.01E+12 &  7.43E+10 &  1.37E+10  & 4.02E+09 &   3.08E+09 & 3.50E+09 &  5.58E+09 \\
CH$_4$     &  5.04E+15 &  9.23E+13 &  2.12E+13  & 5.25E+12 &   1.40E+12 & 1.00E+12 &  5.48E+11 \\
C$_2$H$_2$ &  1.79E+16 &  4.03E+14 &  8.66E+13  & 2.10E+13 &   9.70E+12 & 1.06E+13 &  1.77E+13 \\
C$^+$      &  9.13E+07 &  1.14E+09 &  1.27E+10  & 1.45E+11 &   1.65E+12 & 3.31E+12 &  1.62E+13 \\
CH$_3^+$   &  8.72E+08 &  6.02E+09 &  2.65E+10  & 9.90E+10 &   3.32E+11 & 4.76E+11 &  1.04E+12 \\
C$_2$H$_3^+$  &  1.19E+08 &  2.66E+07 &  3.41E+07  & 6.07E+07 &   1.99E+08 & 3.30E+08 &  6.73E+08 \\
N          &  2.37E+15 &  5.39E+15 &  1.26E+16  & 3.37E+16 &   7.21E+16 & 8.21E+16 &  1.14E+17 \\
N$_2$      &  1.97E+17 &  2.21E+17 &  2.22E+17  & 2.09E+17 &   1.84E+17 & 1.77E+17 &  1.95E+17 \\
NH$_3$     &  5.37E+16 &  7.76E+15 &  1.97E+15  & 2.57E+15 &   3.10E+15 & 3.06E+15 &  3.12E+15 \\
HCN        &  8.35E+15 &  3.58E+15 &  1.27E+15  & 9.11E+14 &   7.92E+14 & 7.69E+14 &  8.42E+14 \\
NH$_3^+$   &  1.08E+08 &  7.65E+08 &  3.77E+09  & 3.14E+10 &   2.63E+11 & 4.50E+11 &  1.34E+12 \\
N$_2$H$^+$ &  7.34E+07 &  1.34E+09 &  1.65E+10  & 1.36E+11 &   7.97E+11 & 1.30E+12 &  3.72E+12 \\
HCNH$^+$   &  4.55E+10 &  1.04E+11 &  6.30E+10  & 6.68E+10 &   1.30E+11 & 1.74E+11 &  5.51E+11 \\
NO         &  4.99E+11 &  3.56E+12 &  2.72E+13  & 1.07E+14 &   4.13E+14 & 6.32E+14 &  1.78E+15 \\
S          &  1.06E+15 &  3.48E+15 &  8.01E+15  & 1.30E+16 &   1.25E+16 & 1.25E+16 &  1.50E+16 \\
S$^+$      &  1.44E+08 &  1.39E+09 &  4.21E+10  & 9.53E+11 &   1.06E+13 & 2.01E+13 &  9.82E+13 \\
SO         &  4.79E+13 &  2.18E+14 &  3.76E+14  & 4.53E+14 &   5.80E+14 & 6.79E+14 &  1.33E+15 \\
SO$_2$     &  8.99E+16 &  9.15E+16 &  9.18E+16  & 9.17E+16 &   9.06E+16 & 8.96E+16 &  1.02E+17 \\
CS         &  6.98E+14 &  6.98E+13 &  4.66E+13  & 5.60E+13 &   8.39E+13 & 9.64E+13 &  1.51E+14 \\
\hline
\multicolumn{8}{l}{$^\dagger$ $\Lx$ in $\ergpers$ and warm columns in $\persqcm$.} \\
\multicolumn{8}{l}{$^*$ Reference case.} \\  
\end{tabular}
\end{center}
\clearpage

\vspace{-0.5in}

\begin{center}
\begin{tabular}{|llllllll|}    
\multicolumn{8}{c}{Table 5. Dependence of Warm Columns (200-2000\,K) on O/C Ratio$^\dagger$} \\
\hline
O/C 	     & 0.25      & 0.50      & 1.00      & 1.50      & 2.00      & 2.5*      & 5.00	\\
\hline
\hline
H$_2$        & 3.85E+21  & 3.88E+21  & 3.88E+21  & 3.91E+21  & 3.91E+21  & 3.91E+21  & 3.93E+21 \\
H$^+$        & 9.86E+13  & 4.40E+14  & 9.21E+14  & 1.09E+13  & 4.86E+12  & 3.20E+12  & 1.44E+12 \\
H$_3^+$      & 3.40E+13  & 4.06E+13  & 5.08E+13  & 3.41E+13  & 2.41E+13  & 1.92E+13  & 1.11E+13 \\
He$^+$       & 2.18E+13  & 2.95E+13  & 1.42E+13  & 1.05E+13  & 8.81E+12  & 7.75E+12  & 5.64E+12 \\
O   	     & 1.61E+12  & 4.85E+12  & 1.17E+17  & 1.66E+17  & 1.76E+17  & 1.70E+17  & 1.45E+17 \\
OH           & 2.78E+11  & 8.09E+11  & 4.31E+13  & 2.67E+14  & 3.93E+14  & 4.57E+14  & 6.82E+14 \\
H$_2$O       & 3.68E+10  & 1.64E+11  & 9.15E+14  & 1.48E+17  & 3.45E+17  & 5.40E+17  & 1.62E+18 \\
H$_3$O$^+$   & 1.26E+08  & 6.45E+08  & 2.28E+12  & 2.78E+13  & 3.68E+13  & 4.22E+13  & 6.24E+13 \\
O$_2$        & 2.21E+08  & 1.50E+09  & 5.74E+13  & 2.42E+16  & 1.95E+17  & 3.72E+17  & 1.25E+18 \\
CO           & 1.79E+16  & 3.69E+16  & 8.78E+17  & 1.07E+18  & 1.08E+18  & 1.09E+18  & 1.11E+18 \\
CO$_2$       & 3.65E+12  & 1.60E+13  & 5.85E+14  & 1.27E+15  & 1.27E+15  & 1.26E+15  & 1.20E+15 \\
HCO          & 2.54E+17  & 5.12E+17  & 7.91E+16  & 3.17E+16  & 1.94E+16  & 1.33E+16  & 3.70E+15 \\
HCO$^+$      & 5.40E+12  & 1.39E+13  & 7.16E+13  & 5.00E+13  & 2.97E+13  & 2.28E+13  & 1.47E+13 \\
C            & 5.45E+17  & 4.25E+17  & 1.12E+17  & 7.23E+14  & 3.83E+14  & 2.68E+14  & 9.01E+13 \\
CH$_3$       & 2.36E+11  & 9.99E+10  & 2.37E+10  & 3.89E+09  & 3.50E+09  & 3.78E+09  & 6.59E+09 \\
CH$_4$       & 1.88E+11  & 1.39E+11  & 3.03E+11  & 1.23E+12  & 1.03E+12  & 9.59E+11  & 7.56E+11 \\
C$_2$H$_2$   & 1.24E+17  & 5.64E+16  & 9.34E+15  & 2.92E+13  & 1.53E+13  & 1.12E+13  & 3.85E+12 \\
C$^+$        & 1.80E+13  & 1.53E+13  & 8.14E+14  & 3.77E+13  & 7.11E+12  & 4.11E+12  & 1.16E+12 \\
CH$_3^+$     & 2.53E+12  & 2.33E+12  & 1.00E+12  & 9.03E+11  & 6.84E+11  & 5.48E+11  & 2.91E+11 \\
C$_2$H$_3^+$ & 1.51E+12  & 8.55E+11  & 3.77E+10  & 1.53E+09  & 6.80E+08  & 3.79E+08  & 5.23E+07 \\
N            & 3.37E+17  & 3.76E+17  & 3.19E+17  & 1.57E+17  & 1.10E+17  & 9.01E+16  & 5.34E+16 \\
N$_2$        & 4.90E+16  & 3.48E+16  & 7.32E+16  & 1.59E+17  & 1.82E+17  & 1.92E+17  & 2.12E+17 \\
NH$_3$       & 4.76E+15  & 5.67E+15  & 1.05E+14  & 1.20E+15  & 2.87E+15  & 3.42E+15  & 3.69E+15 \\
HCN          & 1.93E+16  & 9.55E+15  & 8.13E+15  & 2.10E+15  & 1.17E+15  & 8.05E+14  & 2.90E+14 \\
NH$_3^+$     & 1.17E+12  & 1.45E+12  & 1.96E+12  & 1.28E+12  & 7.56E+11  & 5.30E+11  & 2.35E+11 \\
N$_2$H$^+$   & 4.65E+11  & 3.87E+11  & 4.65E+11  & 2.05E+12  & 1.81E+12  & 1.53E+12  & 9.69E+11 \\
NH$_3^+$     & 1.17E+12  & 1.45E+12  & 1.96E+12  & 1.28E+12  & 7.56E+11  & 5.30E+11  & 2.35E+11 \\
HCNH$^+$     & 1.08E+13  & 9.80E+12  & 2.96E+12  & 4.42E+11  & 2.60E+11  & 1.96E+11  & 1.23E+11 \\
NO           & 8.48E+10  & 3.24E+11  & 3.89E+13  & 4.25E+14  & 6.28E+14  & 7.29E+14  & 1.05E+15 \\
S            & 1.01E+17  & 1.03E+17  & 9.06E+16  & 1.51E+16  & 1.36E+16  & 1.28E+16  & 1.16E+16 \\
S$^+$        & 5.90E+15  & 6.06E+15  & 9.44E+15  & 6.26E+13  & 3.04E+13  & 2.42E+13  & 1.83E+13 \\
SO           & 5.94E+10  & 2.35E+11  & 9.94E+13  & 4.32E+14  & 5.78E+14  & 7.23E+14  & 1.31E+15 \\
SO$_2$       & 3.54E+07  & 3.96E+08  & 1.04E+16  & 9.59E+16  & 9.72E+16  & 9.79E+16  & 9.94E+16 \\
CS           & 2.55E+15  & 1.44E+15  & 1.55E+14  & 9.78E+13  & 1.07E+14  & 1.01E+14  & 5.20E+13 \\

\hline
\multicolumn{8}{l}{$^\dagger$ Warm columns in $\persqcm$.} \\
\multicolumn{2}{l}{$^*$ Reference case.}\\
\end{tabular}
\end{center}
\clearpage

\begin{center}
\begin{tabular}{|lllllll|}    
\multicolumn{7}{c}{Table 6. Dependence of Warm Columns (200-2000\,K) on Disk Radius$^\dagger$} \\
\hline
R (AU) 	     & 0.25      & 0.5      & 1*      & 2      & 4      & 10     \\
\hline
\hline
H$_2$        & 4.85E+23 & 1.54E+22 & 3.91E+21 & 2.35E+21 & 1.68E+21 & 7.76E+20 \\
H$^+$        & 9.89E+12 & 4.81E+12 & 3.20E+12 & 2.29E+12 & 1.74E+12 & 2.40E+12 \\
H$_3^+$      & 3.68E+13 & 2.63E+13 & 1.92E+13 & 1.49E+13 & 1.16E+13 & 9.47E+12 \\
O            & 1.59E+16 & 4.45E+17 & 1.70E+17 & 9.61E+16 & 7.29E+16 & 7.95E+16 \\
OH           & 8.01E+14 & 8.10E+14 & 4.57E+14 & 3.34E+14 & 2.94E+14 & 4.20E+14 \\
H$_2$O       & 1.66E+20 & 9.75E+17 & 5.40E+17 & 4.17E+17 & 3.00E+17 & 5.40E+16 \\
H$_3$O$^+$   & 1.30E+14 & 6.21E+13 & 4.22E+13 & 3.20E+13 & 2.26E+13 & 6.92E+12 \\
O$_2$        & 6.54E+18 & 2.08E+18 & 3.72E+17 & 1.85E+17 & 1.28E+17 & 7.81E+16 \\
CO           & 1.30E+20 & 4.28E+18 & 1.09E+18 & 6.56E+17 & 4.74E+17 & 2.25E+17 \\
CO$_2$       & 1.30E+16 & 1.77E+15 & 1.26E+15 & 1.09E+15 & 7.73E+14 & 2.22E+13 \\
HCO          & 3.21E+16 & 1.72E+16 & 1.33E+16 & 1.11E+16 & 5.62E+15 & 9.52E+11 \\
HCO$^+$      & 1.89E+13 & 3.42E+13 & 2.28E+13 & 1.67E+13 & 1.41E+13 & 1.82E+13 \\
C            & 6.60E+14 & 3.61E+14 & 2.68E+14 & 2.44E+14 & 2.36E+14 & 3.77E+12 \\
CH$_3$       & 1.20E+10 & 5.39E+09 & 3.78E+09 & 2.65E+09 & 1.38E+09 & 1.50E+08 \\
CH$_4$       & 2.33E+16 & 4.33E+12 & 9.59E+11 & 6.70E+11 & 4.73E+11 & 4.34E+10 \\
C$_2$H$_2$   & 9.11E+17 & 1.59E+13 & 1.12E+13 & 9.63E+12 & 6.79E+12 & 2.67E+10 \\
C$^+$        & 5.35E+12 & 4.89E+12 & 4.11E+12 & 3.43E+12 & 3.26E+12 & 5.42E+12 \\
CH$_3^+$     & 1.44E+12 & 7.73E+11 & 5.48E+11 & 3.76E+11 & 1.68E+11 & 1.56E+09 \\
C$_2$H$_3^+$ & 9.87E+08 & 5.13E+08 & 3.79E+08 & 3.31E+08 & 2.48E+08 & 8.05E+05 \\
N            & 2.93E+17 & 1.67E+17 & 9.01E+16 & 6.37E+16 & 4.42E+16 & 1.58E+16 \\
N$_2$        & 2.50E+19 & 8.37E+17 & 1.92E+17 & 1.11E+17 & 8.08E+16 & 4.00E+16 \\
NH$_3$       & 4.77E+18 & 1.29E+16 & 3.42E+15 & 1.76E+15 & 1.22E+15 & 6.51E+14 \\
HCN          & 3.12E+18 & 2.50E+15 & 8.05E+14 & 4.99E+14 & 3.51E+14 & 1.34E+14 \\
NH$_3^+$     & 1.22E+12 & 7.20E+11 & 5.30E+11 & 4.18E+11 & 3.00E+11 & 1.30E+11 \\
N$_2$H$^+$   & 1.89E+12 & 2.25E+12 & 1.53E+12 & 1.13E+12 & 9.42E+11 & 1.14E+12 \\
HCNH$^+$     & 1.20E+12 & 2.93E+11 & 1.96E+11 & 1.45E+11 & 9.66E+10 & 3.83E+10 \\
NO           & 1.23E+15 & 1.31E+15 & 7.29E+14 & 5.32E+14 & 4.72E+14 & 6.63E+14 \\
S            & 3.10E+16 & 1.69E+16 & 1.28E+16 & 9.75E+15 & 4.47E+15 & 1.07E+12 \\
S$^+$        & 5.99E+13 & 3.32E+13 & 2.42E+13 & 2.13E+13 & 1.93E+13 & 2.59E+11 \\
SO           & 1.91E+15 & 9.82E+14 & 7.23E+14 & 6.10E+14 & 4.29E+14 & 1.39E+13 \\
SO$_2$       & 1.36E+19 & 4.15E+17 & 9.79E+16 & 5.70E+16 & 4.35E+16 & 2.27E+16 \\
CS           & 2.65E+14 & 1.33E+14 & 1.01E+14 & 9.06E+13 & 7.26E+13 & 6.11E+09 \\
\hline
\multicolumn{7}{l}{$^\dagger$ Warm columns in $\persqcm$.} \\
\multicolumn{2}{l}{$^*$ Reference case.}\\
\end{tabular}
\end{center}
\clearpage

\begin{figure}
\plotfiddle{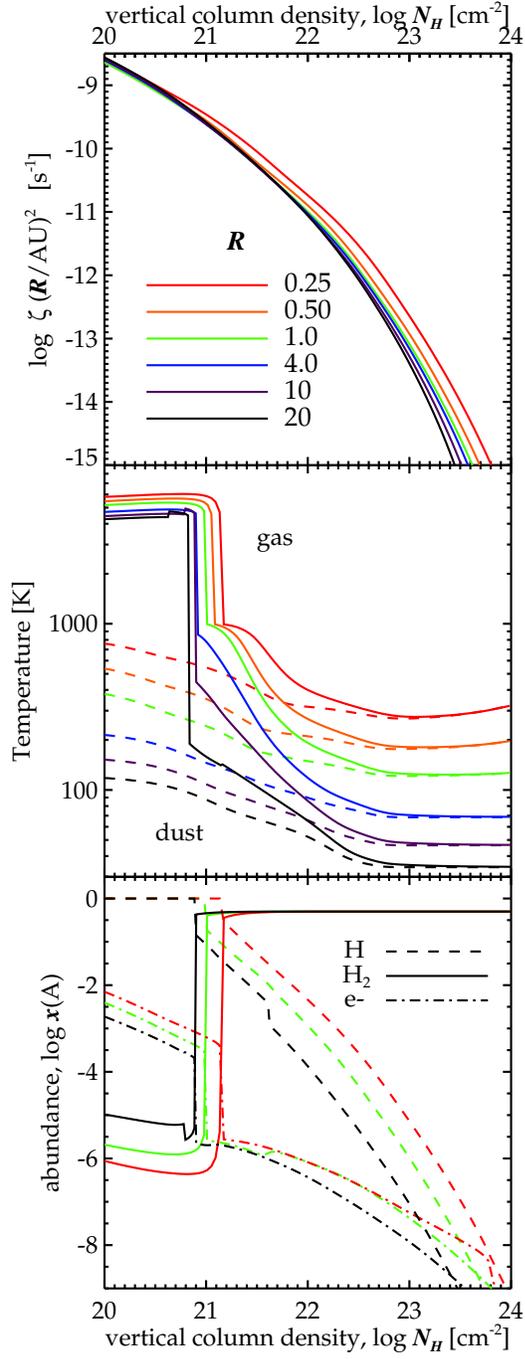}{7.0truein}{0}{70}{70}{-100}{-10}
\caption{X-ray ionization rate (top panel), gas and dust temperature 
(middle panel), and abundances of atomic hydrogen, molecular 
hydrogen, and electrons (bottom panel) as a function of vertical 
column density and radius in the reference model. 
\label{figa}}
\end{figure}
\clearpage

\begin{figure}
\plotfiddle{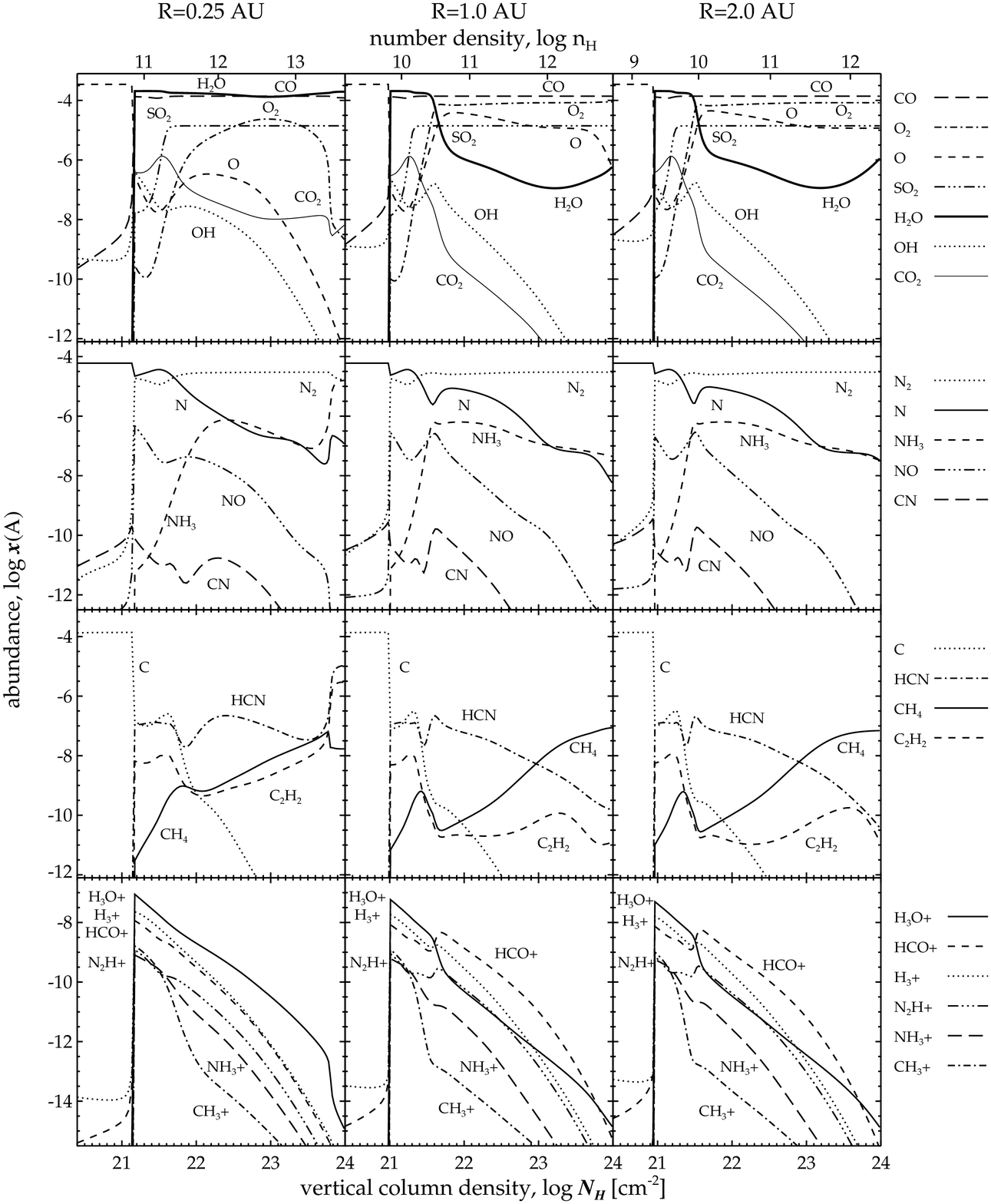}{8.25truein}{0}{60}{60}{-120}{0}
\caption{(a) Abundances of various species in the reference model at 
0.25\,AU (left), 1\,AU (middle), and 2\,AU (right).
\label{figb1}}
\end{figure}
\clearpage

\addtocounter{figure}{-1}

\begin{figure}
\plotfiddle{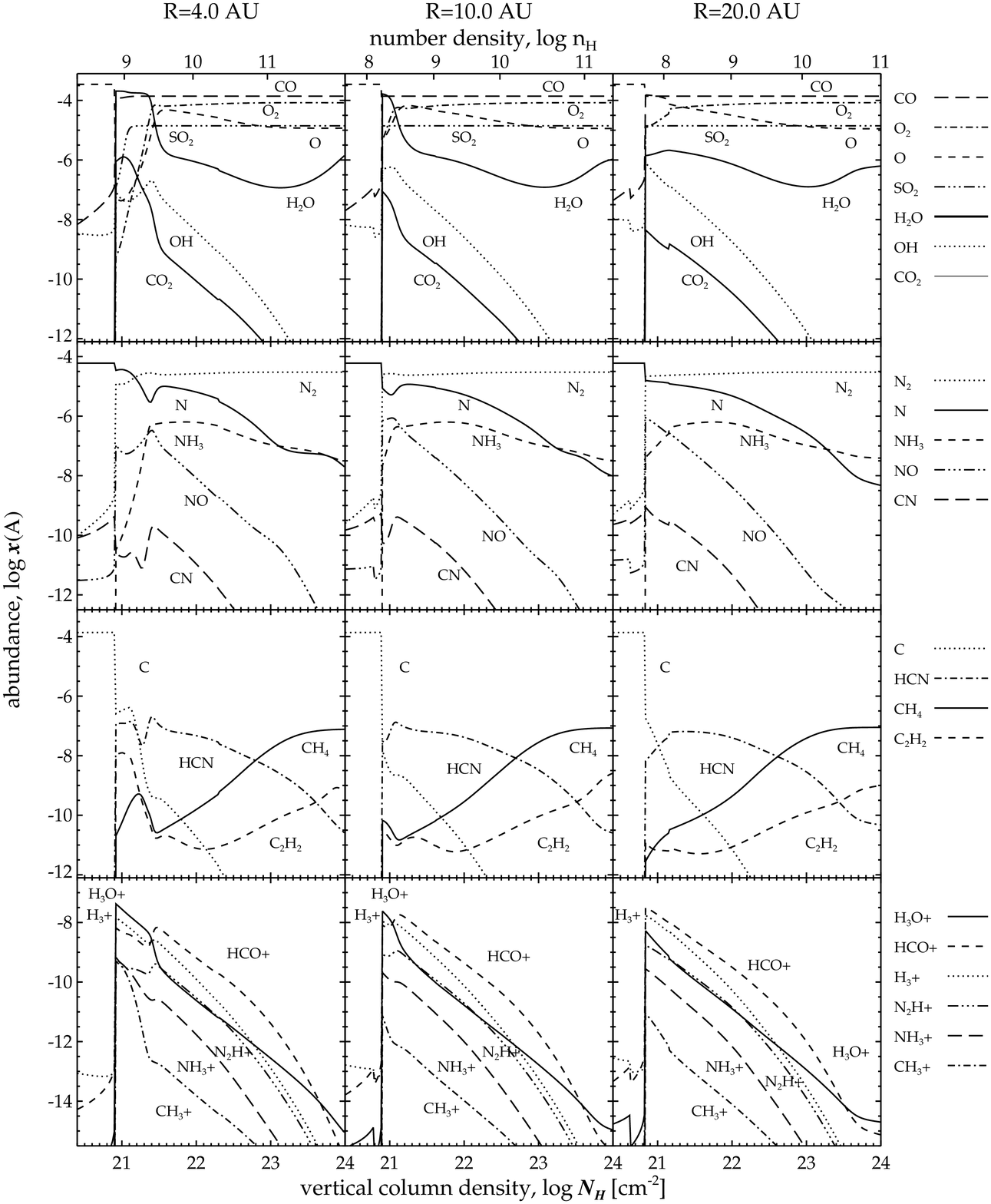}{8.25truein}{0}{60}{60}{-120}{0}
\caption{(b) Abundances of various species in the reference model at 
4\,AU (left), 10\,AU (middle), and 20\,AU (right).
\label{figb2}}
\end{figure}
\clearpage

\begin{figure}
\plotfiddle{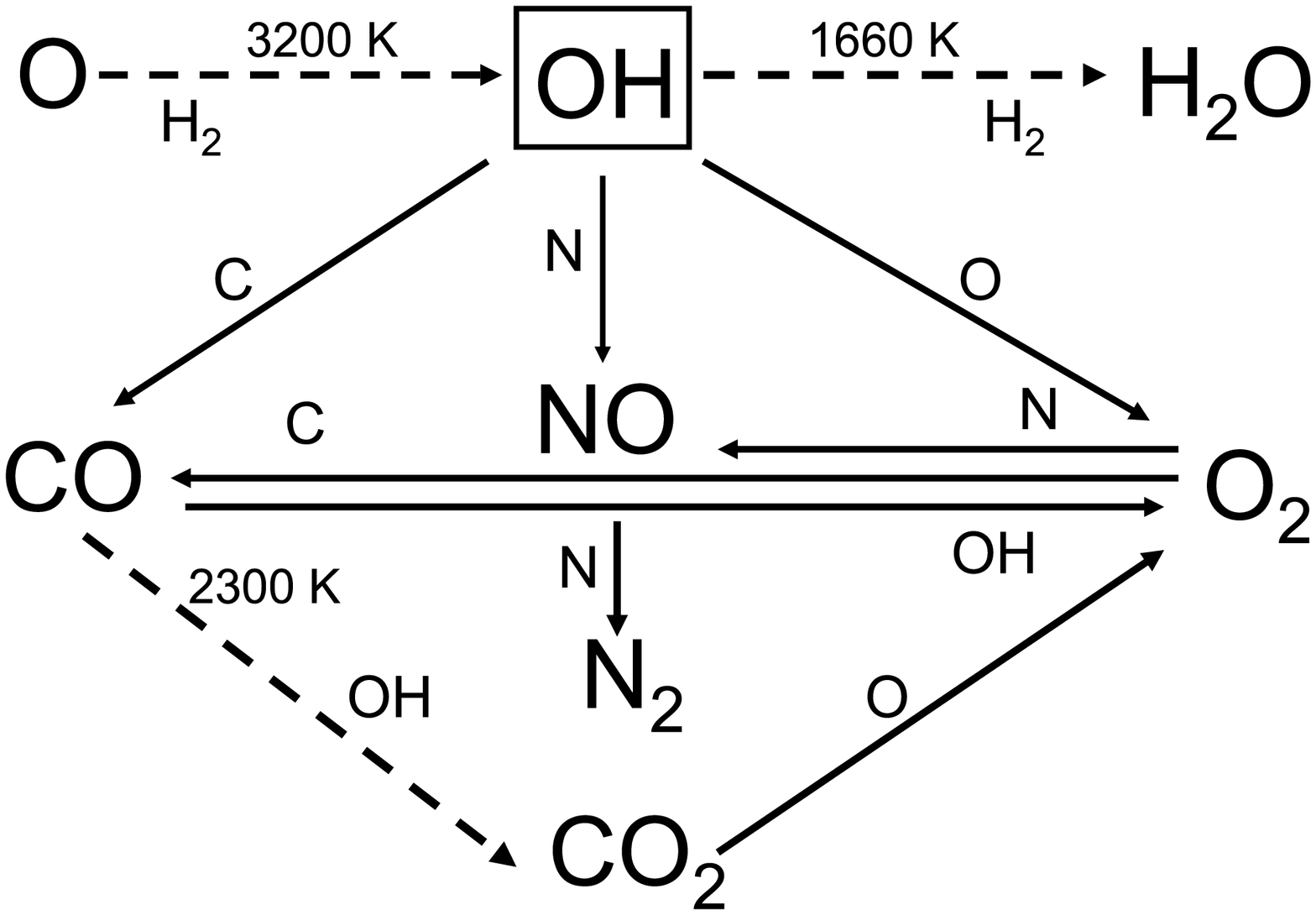}{5.0truein}{0}{70}{70}{-200}{-30}
\caption{Schematic chemical network showing how fast neutral reactions
of OH with C, O, and N atoms generate oxides that provide pathways
to $\COtwo$, $\Otwo$, and $\Ntwo$.  
\label{fig3}}
\end{figure}
\clearpage

\begin{figure}
\plotfiddle{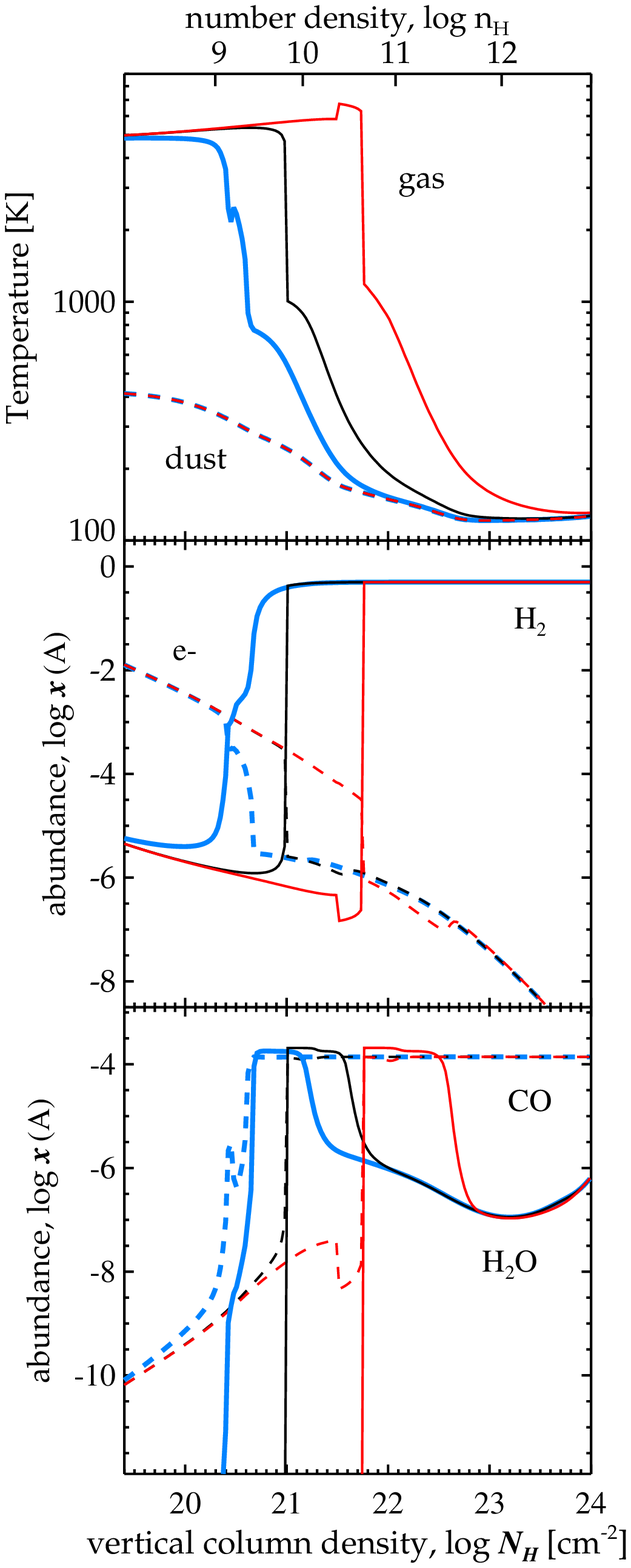}{7.25truein}{0}{70}{70}{-130}{0}
\caption{Gas (solid) and dust (dashed) temperatures (top panel), 
electron (dashed) and $\Htwo$ (solid) abundances (middle panel), and 
CO (dashed) and water (solid) abundances (bottom panel)
in the reference case (black), 
and for 
$\alpha_h=1$, $a_g=7.07$ (red) and 
$\alpha_h=0.01$, $a_g=0.707$ (blue). 
\label{fig4}}
\end{figure}

\begin{figure}
\plotfiddle{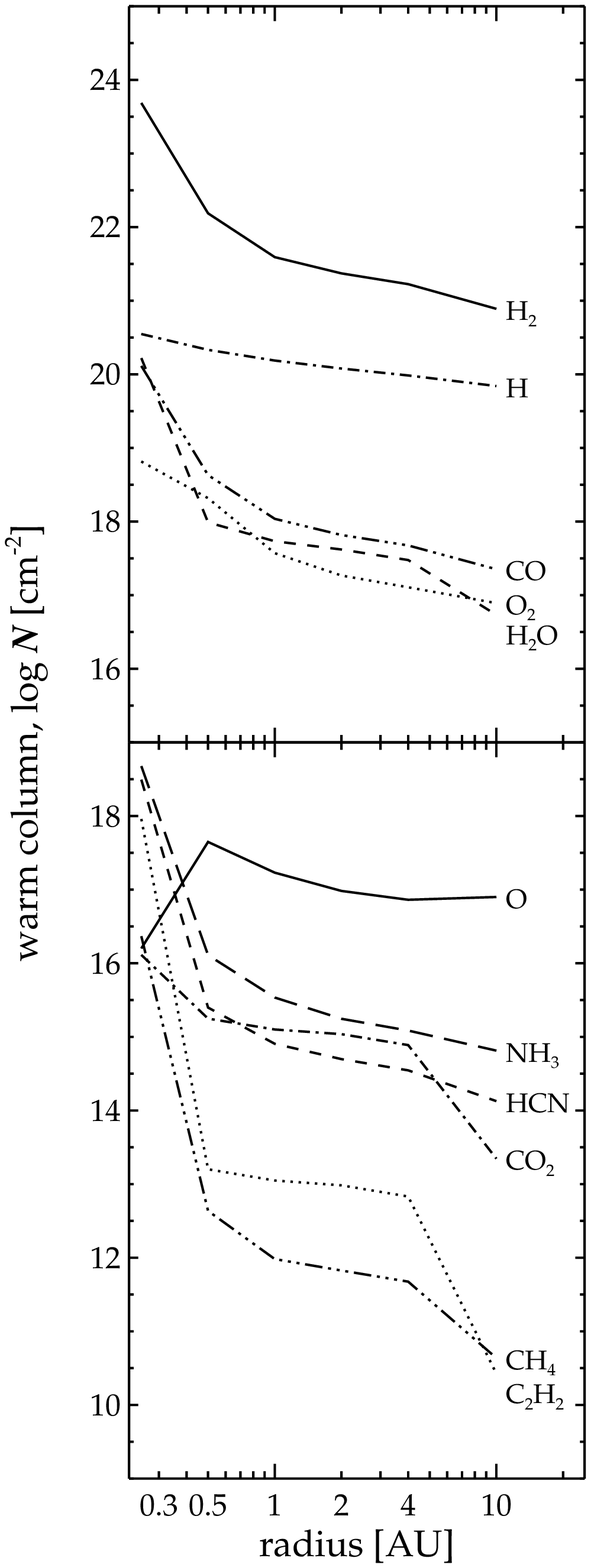}{8.0truein}{0}{75}{75}{-125}{0}
\caption{Warm columns of various species as a function of disk radius in 
the reference case. 
\label{fig5}}
\end{figure}

\clearpage

\end{document}